\documentclass[12pt]{article}

\usepackage[dvips]{graphicx}
\usepackage{epsfig}
\usepackage{colordvi}
\usepackage{anysize}
\usepackage{amsmath}
\usepackage{amssymb}
\usepackage{amscd}
\usepackage{fancyhdr}
\usepackage{hyperref}
\usepackage{amsthm}

\begin{document}

%%%%%%%%%%%%%%%%%%%%%%%%
%
% Sheer's Standard Custom Commands
%
%%%%%%%%%%%%%%%%%%%%%%%%

%%%%%%
%
%  Characters and Scripts
%
%%%%%%

%
% Greek
%
\newcommand{\al}{\alpha}
\newcommand{\bt}{\beta}
\newcommand{\gm}{\gamma}
\newcommand{\Gm}{\Gamma}
\newcommand{\dl}{\delta}
\newcommand{\Dl}{\Delta}
\newcommand{\ep}{\epsilon}
\newcommand{\varep}{\varepsilon}
\newcommand{\zt}{\zeta}
\newcommand{\et}{\eta}
\newcommand{\tht}{\theta}
\newcommand{\vth}{\vartheta}
\newcommand{\io}{\iota}
\newcommand{\kp}{\kappa}
\newcommand{\lm}{\lambda}
\newcommand{\Lm}{\Lambda}
\newcommand{\rh}{\rho}
\newcommand{\sg}{\sigma}
\newcommand{\Sg}{\Sigma}
\newcommand{\ta}{\tau}
\newcommand{\ph}{\phi}
\newcommand{\Ph}{\Phi}
\newcommand{\phv}{\varphi}
\newcommand{\ch}{\chi}
\newcommand{\ps}{\psi}
\newcommand{\Ps}{\Psi}
\newcommand{\om}{\omega}
\newcommand{\Om}{\Omega}

%
% Dotted Greek
%
\newcommand{\dal}{\dot{\alpha}}
\newcommand{\dbt}{\dot{\beta}}
\newcommand{\dgm}{\dot{\gamma}}
\newcommand{\ddl}{\dot{\delta}}

%%
%% blackboard
%%
\newcommand{\A}{\mathbb{A}}
\newcommand{\C}{\mathbb{C}}
\newcommand{\B}{\mathbb{B}}
\newcommand{\D}{\mathbb{D}}
\newcommand{\BP}{\mathbb{P}}
\newcommand{\G}{\mathbb{G}}
\newcommand{\I}{\mathbb{I}}
\newcommand{\K}{\mathbb{K}}
\newcommand{\N}{\mathbb{N}}
\newcommand{\Q}{\mathbb{Q}}
\newcommand{\R}{\mathbb{R}}
\newcommand{\T}{\mathbb{T}}
\newcommand{\Z}{\mathbb{Z}}

%
% Calligraphic
%
\newcommand{\CA}{\mathcal{A}}
\newcommand{\CB}{\mathcal{B}}
\newcommand{\CC}{\mathcal{C}}
\newcommand{\F}{\mathcal{F}}
\newcommand{\CalD}{\mathcal{D}}
\newcommand{\CE}{\mathcal{E}}
\newcommand{\CG}{\mathcal{G}}
\renewcommand{\H}{\mathcal{H}}
\newcommand{\CJ}{\mathcal{J}}
\newcommand{\CI}{\mathcal{I}}
\newcommand{\CK}{\mathcal{K}}
\newcommand{\CL}{\mathcal{L}}
\newcommand{\CM}{\mathcal{M}}
\newcommand{\CN}{\mathcal{N}}
\newcommand{\CS}{\mathcal{S}}
\newcommand{\CO}{\mathcal{O}}
\newcommand{\CP}{\mathcal{P}}
\newcommand{\CQ}{\mathcal{Q}}
\newcommand{\CR}{\mathcal{R}}
\newcommand{\CT}{\mathcal{T}}
\newcommand{\CV}{\mathcal{V}}
\newcommand{\CW}{\mathcal{W}}
\newcommand{\CU}{\mathcal{U}}
\newcommand{\CY}{\mathcal{Y}}
\newcommand{\CZ}{\mathcal{Z}}
\renewcommand{\P}{\mathcal{P}}
\renewcommand{\L}{\mathcal{L}}

%
% Frankish
%
\newcommand{\FG}{\mathfrak{G}}
\newcommand{\FU}{\mathfrak{U}}
\newcommand{\Fg}{\mathfrak{g}}
\newcommand{\Fu}{\mathfrak{u}}
\newcommand{\Fh}{\mathfrak{h}}
\newcommand{\Fs}{\mathfrak{s}}

%%%%
%
% Abbreviated Common Commands
%
%%%

%
% Arrows and commands for Exact Sequences
%
\newcommand{\onra}[1]{\stackrel{#1}{\ra}}
\newcommand{\onlra}[1]{\stackrel{#1}{\lra}}
\newcommand{\ondbarr}[2]{\mathop{\llra}_{#2}^{#1}}
\newcommand{\ondarr}[1]{\stackrel{#1}{\dra}}
\newcommand{\exact}[3]{0 \lra #1 \lra #2 \lra #3 \lra 0 }
\newcommand{\sexact}[3]{0 \lra #1 \lra #2 \lra #3 \llra 0 }
\newcommand{\SES}[3]{#1 \lra #2 \lra #3}
\newcommand{\LES}[6]{#1 \lra #2 \lra #3 \lra #4 \lra #5 \lra #6 }
\newcommand{\Mexact}[5] { 0 \lra #1 \onlra{#2} #3 \onlra{#4} #5 \lra 0}
\newcommand{\sMexact}[6] { 0 \lra #1 \onlra{#2} #3 \ondbarr{#4}{#5} #6 \lra 0}
\newcommand{\dsMexact}[7]
        {0 \lra #1 \ondbarr{#2}{#3} #4 \ondbarr{#5}{#6} #7 \lra 0}
\newcommand{\MSES}[5] { #1 \onlra{#2} #3 \onlra{#4} #5}
\newcommand{\homot}{\stackrel{h}{\sim}}
\newcommand{\enp}{\qed\\[5pt]}

% Abbreviated commands
\newcommand{\ra}{\rightarrow}
\newcommand{\rla}{\leftrightarrow}
\newcommand{\lra}{\longrightarrow}
\newcommand{\llra}{\longleftrightarrow}
\newcommand{\dra}{\dashrightarrow}
\newcommand{\opl}{\oplus}
\newcommand{\otm}{\otimes}
\newcommand{\sub}{\subset}
\newcommand{\bcap}{\bigcap}
\newcommand{\bcup}{\bigcup}

%
%  Abbreviated Environments
%
\newcommand{\eqnb}{\begin{equation}}
\newcommand{\eqne}{\end{equation}}

%
% New math text
%
\newcommand{\krn}{\textrm{Ker}}
\newcommand{\img}{\textrm{Im}}
\newcommand{\Id}{\textrm{id}}
\newcommand{\Sp}{\textrm{sp}}
\newcommand{\tr}{\textrm{Tr}}
\newcommand{\dtr}{\textrm{det}}

%
% Formatting
%
\newcommand{\und}{\underline}
\newcommand{\ovr}{\overline}
\newcommand{\tld}{\tilde}
\newcommand{\wtld}{\widetilde}
\newcommand{\phnt}{\phantom}

%
% Common differential operators
%
\newcommand{\prt}{\partial}
\newcommand{\prtb}{\overline{\partial}}
\newcommand{\pmu}{\partial^\mu}
\newcommand{\pmd}{\partial_\mu}
\newcommand{\pnu}{\partial^\nu}
\newcommand{\pnd}{\partial_\nu}
\newcommand{\nab}{\nabla}
\newcommand{\nmu}{\nabla^\mu}
\newcommand{\nmd}{\nabla_\mu}
\newcommand{\nnu}{\nabla^\nu}
\newcommand{\nnd}{\nabla_\nu}

%%%%%
%
% QFT
%
%%%%

% Dirac slash
\newcommand{\sh}[1]{#1\hskip-7pt \diagup}
\newcommand{\slsh}[1]{#1\hskip-7pt \slash}
\newcommand{\Sh}[1]{#1\hskip-11pt \diagup}
\newcommand{\SSh}[1]{#1\hskip-13pt \diagup}
\newcommand{\SSSh}[1]{#1\hskip-17pt \diagup}
\newcommand{\Shnine}[1]{#1\hskip-9pt \diagup}
% Spinors
\newcommand{\psidag}{\psi^\dagger}
\newcommand{\psib}{\overline{\psi}}
\newcommand{\psibp}{\overline{\psi}_+}
\newcommand{\psibn}{\overline{\psi}_-}
\newcommand{\psip}{\psi_+}
\newcommand{\psin}{\psi_-}
% Dirac Operator
\newcommand{\dirac}{\sh{\prt} - \sh{A}}
\newcommand{\dirop}{i\Shnine{D}}
\newcommand{\diropnoi}{\Shnine{D}}
\newcommand{\dirnam}{i\slsh{\nabla}}

% d-dimensional integration
\newcommand{\ddint}{\int d^dx\, }
% fermion path integral
\newcommand{\fint}{\int \CalD\psib \CalD\psi\, }

% d-dimensional gamma matrix
\newcommand{\gmd}{\gamma^d}
\newcommand{\gmmu}{\gm^\mu}
\newcommand{\chiralcur}{J^{d\mu}}

%
% Quantum Mechanics
%
%
\newcommand{\bra}[1]{\ensuremath{ \left\langle #1 \right| } }  % quantum state %
\newcommand{\ket}[1]{\ensuremath{ \left| #1 \right\rangle} }  % conjugate quantum state %
\newcommand{\braket}[1]{\ensuremath{\left\langle #1 \right\rangle} }  % conjugate quantum state %

%
%
% SUSY
%
\newcommand{\sX}{ {\mathbf X}}
\newcommand{\sD}{D_{\tht}}
\newcommand{\sDb}{D_{\ovr{\tht}}}
\newcommand{\zb}{\ovr{z}}
\newcommand{\thtb}{\ovr{\tht}}

%
% Common forms
%

% These are generally not needed -- just use dx, dy, dz directly
%\newcommand{\dx}{\textrm{dx}}
%\newcommand{\dy}{\textrm{dy}}
%\newcommand{\dz}{\textrm{dz}}

%%%%%%%
%
%  Other stuff
%
%%%%%%%

\newcounter{enumlistcount}
\newcounter{smallenumcount}
\newcounter{listicounter}
\newcounter{listiicounter}
%
% The following lines force the references to these counters to be of the
% correct form (see environments below to see how we're using these counters)
%
\renewcommand{\thelisticounter}{\alph{listicounter}}
%\renewcommand{\p@listicounter}{\thelisticounter.}

% note that the \usecounter has to be in the size argument, not the {label}.
% Latex is just a weirdo.
\newenvironment{smallenum}{ \begin{list}{\arabic{smallenumcount}.}{\usecounter{smallenumcount}\setlength{\itemsep}{-1pt}\setlength{\leftmargin}{20pt}} } {\end{list}}
\newenvironment{enumlist}{ \begin{list}{\arabic{enumlistcount}.}{\usecounter{enumlistcount}\setlength{\itemsep}{-1pt}\setlength{\leftmargin}{10pt}} } {\end{list}}
\newenvironment{listi}{\begin{list}{\alph{listicounter}.}{\usecounter{listicounter}\setlength{\itemsep}{-1pt}} } {\end{list}}
\newenvironment{listii}{\begin{list}{\arabic{listiicounter}.}{\usecounter{listiicounter}\setlength{\itemsep}{-1pt}} } {\end{list}}

\newtheorem{theorem}{Theorem}[section]
\newtheorem{lemma}[theorem]{Lemma}
\newtheorem{corollary}[theorem]{Corollary}
\newtheorem{definition}[theorem]{Definition}
\newtheorem{proposition}[theorem]{Proposition}

%
% Custom Environment
%
%

\newenvironment{proof2}[1][Proof]{\begin{trivlist} \item[\hskip \labelsep
{\bfseries #1}]}{\end{trivlist}}

\newenvironment{question}{\vspace{0.4cm}\noindent\begin{tabular}{|p{6.1in}|}\hline\vspace{0.1cm }}{ \vspace{0.1cm} \\\hline\end{tabular}\vspace{0.4cm}}

\newenvironment{spequ}{\begin{equation}\begin{split} }{ \end{split}\end{equation}}

\newenvironment{margline}{\noindent\begin{tabular}{|p{6.1in}}}{\end{tabular}}

%
% Customized Margin input
%
%

\newcommand{\cmargin}[1]{\marginpar{%
      \vskip-\baselineskip
      \raggedright\footnotesize
      \itshape\hrule\smallskip#1\par\smallskip\hrule}}

%\hline\rule{1ex}{1ex}
%\hspace{\stretch{1}}}

%{\hspace{\stretch{1}}%
%\rule{1ex}{1ex}\hline\end{tabular}}

%%%%%%%%%%%%%%%%%%%%%%%%
%
% End: Sheer's Custom Commands
%
%%%%%%%%%%%%%%%%%%%%%%%%

%%%%%%%%%%%%%%%%%%%%%%%%
%
% File-Specific Custom Commands
%
%%%%%%%%%%%%%%%%%%%%%%%%

%\newcommand{\and}{\nu(\phi_1)}
\newcommand{\anu}{\nu(\phi_1)}
\newcommand{\bnd}{\nu(\phi_2)}
\newcommand{\bnu}{\nu(\phi_2)}
\newcommand{\amd}{\mu(\phi_1)}
\newcommand{\amu}{\mu(\phi_1)}
\newcommand{\bmd}{\mu(\phi_2)}
\newcommand{\bmu}{\mu(\phi_2)}
\newcommand{\phand}{\phi_{1 \nu}}
\newcommand{\phanu}{\phi^\nu_1}
\newcommand{\phbnd}{\phi_{2 \nu}}
\newcommand{\phbnu}{\phi^\nu_2}
\newcommand{\phamd}{\phi_{1 \mu}}
\newcommand{\phamu}{\phi^\mu_1}
\newcommand{\phbmd}{\phi_{2 \mu}}
\newcommand{\phbmu}{\phi^\mu_2}

\newcommand{\upar}{\uparrow}
\newcommand{\dnar}{\downarrow}

\newcommand{\Qd}{Q^\dag}

\newcommand{\Gns}{G_{-\frac{1}{2}}}
\newcommand{\Gdn}{G_{-\frac{1}{2}}^\downarrow}
\newcommand{\Gup}{G_{-\frac{1}{2}}^\uparrow}

\newcommand{\Gzns}{G_{0}}
\newcommand{\Gzdn}{G_{0}^\downarrow}
\newcommand{\Gzup}{G_{0}^\uparrow}

\newcommand{\cgh}{\bar{c}\;}
\newcommand{\tA}{\tld{A}}
\newcommand{\ttht}{\tld{\tht}}

\newcommand{\hrho}{\hat{\rho}}

%%%%%%%%%%%%%%%%%%%%%%%%
%
% End Custom Commands
%
%%%%%%%%%%%%%%%%%%%%%%%%

\pagestyle{empty}
\begin{flushright}
\begin{tabular}{ll}
DAMTP-2006-99 & \\
EMPG-06-10 & \\
ITFA-06-41 & \\
MCTP-06-28 & \\
hep-th/0611080 & \\
07/11/06 & \\ [.3in]
\end{tabular}
\end{flushright}
\begin{center}
{\Large {\bf{Open $G_2$ Strings}}} \\ [.47in]
{Jan de Boer$\,{}^{1}$, {Paul de Medeiros$\, {}^{2}$, Sheer El-Showk$\,{}^{1}$
and Annamaria Sinkovics$\,{}^{3}$}}
\\ [.27in]
$\,{}^{1}$\, {\emph{Instituut voor Theoretische Fysica, Valckenierstraat 65,\\
1018 XE Amsterdam, The Netherlands.}} \\ [.27in]
$\,{}^{2}$\, {\emph{School of Mathematics and Maxwell Institute for Mathematical Sciences, \\
University of Edinburgh, King's Buildings, Mayfield Road, \\
Edinburgh EH9 3JZ, Scotland, U.K.}} \\ [.27in]
$\,{}^{3}$\, {\emph{Department of Applied Mathematics and Theoretical Physics, \\
Centre for Mathematical Sciences, University of Cambridge, \\
Wilberforce Road, Cambridge, CB3 0WA, U.K.}} \\ [.27in]
{\tt{jdeboer@science.uva.nl}}, {\tt{p.demedeiros@ed.ac.uk}}, \\
{\tt{sheer@science.uva.nl}}, {\tt{a.sinkovics@damtp.cam.ac.uk}}
\\ [.45in]

{\large{\bf{Abstract}}} \\ [.2in]
\end{center}
We consider an open string version of the topological twist previously proposed
for sigma-models with $G_2$ target spaces. We determine the cohomology of open
strings states and relate these to geometric deformations of calibrated
submanifolds and to flat or anti-self-dual connections on such submanifolds. On
associative three-cycles we show that the worldvolume theory is a gauge-fixed
Chern-Simons theory coupled to normal deformations of the cycle.  For
coassociative four-cycles we find a functional that extremizes on anti-self-dual
gauge fields. A brane wrapping the whole $G_2$ induces a seven-dimensional
associative Chern-Simons theory on the manifold.  This theory has already been
proposed by Donaldson and Thomas as the higher-dimensional generalization of
real Chern-Simons theory.  When the $G_2$ manifold has the structure of a
Calabi-Yau times a circle, these theories reduce to a combination of the open
A-model on special Lagrangians and the open B+${\bar {\mbox{B}}}$-model on
holomorphic submanifolds. We also comment on possible applications of our
results.

\clearpage
\pagestyle{plain}
\pagenumbering{arabic}

%\doublespacing
%\onehalfspacing

%%%%%
%
% Document Starts here
%
%%%%%

\tableofcontents
\newpage

%%%%%%%%%%%%%%%%%%%%
%
% Main Text
%
%%%%%%%%%%%%%%%%%%%%

\section{Introduction}

Topological strings have been studied quite intensively as a toy
model of ordinary string theory. Besides displaying a rich
mathematical structure, they partially or completely control
certain BPS quantities in ordinary string theory, and as such have
found applications e.g. in the study BPS black holes and
non-perturbative contributions to superpotentials.

Unfortunately, a full non-perturbative definition of topological
string theory is still lacking, but it is clear that it will
involve ingredients from both the A- and B-model, and that both
open and closed topological strings will play a role. Since
M-theory is crucial in understanding the strong coupling limit and
nonperturbative properties of string theory, one may wonder
whether something similar is true in the topological case, i.e.
does there exist a seven-dimensional topological theory which
reduces to topological string theory in six dimensions when
compactified on a circle? And could such a seven-dimensional
theory shed light on the non-perturbative properties of
topological string theory?

In order to find such a seven-dimensional theory one can use
various strategies. One can try to directly guess the spacetime
theory, as in \cite{Dijkgraaf:2004te,Nekrasov:2005bb}, or one can
try to construct a topological membrane theory as in
\cite{Grassi:2004xr,Bao:2005yx,Anguelova:2005cv,Bonelli:2005rw,Ikeda:2006pd}
(after all, M-theory appears to be a theory of membranes, though
the precise meaning of this sentence remains opaque). In this
paper we will follow a different approach and study a topological
version of strings propagating on a manifold of $G_2$ holonomy,
following \cite{deBoer:2005pt} (for an earlier work on $G_2$
sigma-models see \cite{Shatashvili:1994zw}). In
\cite{deBoer:2005pt} the topological twist was defined using the
extended worldsheet algebra that sigma-models on manifolds with
exceptional holonomy possess \cite{Howe:1991ic}. For manifolds of
$G_2$ holonomy the extended worldsheet algebra contains the
$c=7/10$ superconformal algebra \cite{Shatashvili:1994zw} that
describes the tricritical Ising model, and the conformal block
structure of this theory was crucial in defining the twist. In
\cite{deBoer:2005pt} it was furthermore shown that the BRST
cohomology of the topological $G_2$ string is equivalent to the
ordinary de Rham cohomology of the seven-manifold, and that the
genus zero three-point functions are the third derivatives of a
suitable prepotential, which turned out to be equal to the
seven-dimensional Hitchin functional of \cite{Hitchin:2000jd}. The
latter also features prominently in
\cite{Dijkgraaf:2004te,Nekrasov:2005bb}, suggesting a close
connection between the spacetime and worldsheet approaches.

In the present paper we will study open topological strings on
seven-manifolds of $G_2$ holonomy, using the same twist as in
\cite{deBoer:2005pt}. There are several motivations to do this.
First of all, we hope that this formalism will eventually lead to
a better understanding of the open topological string in six
dimensions. Second, some of the results may be relevant for the
study of realistic compactifications of M-theory on manifolds of
$G_2$ holonomy\footnote{This will require an extension of our
results to singular manifolds which is an interesting direction
for future research.}, for a recent discussion of the latter see
e.g. \cite{Acharya:2006ia}. Third, by studying branes wrapping
three-cycles we may establish a connection between topological
strings and topological membranes in seven dimensions. And
finally, for open topological strings one can completely determine
the corresponding open string field theory \cite{Witten:1992fb},
from which one can compute arbitrary higher genus partition
functions and from which one can also extract highly non-trivial
all-order results for the closed topological string using
geometric transitions \cite{Gopakumar:1998vy}. Repeating such an
analysis in the $G_2$ case would allow us to use open $G_2$ string
field theory to perform computations at higher genus in both the
open and closed topological $G_2$ string. This is of special
importance since the definition and existence of the topological
twist at higher genus has not yet been rigorously established in
the $G_2$ case.

Along the way we will run into various interesting mathematical
structures and topological field theories in various dimensions
that may be of interest in their own right.

The outline and summary of this paper is as follows. We will first
briefly review the closed topological $G_2$ string and its Hilbert
space. We will then consider open topological strings and their
boundary conditions. Consistent boundary conditions are those
which preserve one copy of the non-linear $G_2$ worldsheet algebra
and were previously analyzed in \cite{Becker:1996ay,Howe:2005je}.
One finds that there are topological zero-, three-, four- and
seven-branes in the theory\footnote{It is unclear to us how we
could incorporate coisotropic six-branes in our theory, whose
existence is suggested in \cite{Bao:2006ef}.}. The three- and
four-branes wrap associative and coassociative cycles respectively
and are calibrated by the covariantly constant three-form and its
Hodge-dual which define the $G_2$ structure.

Next, we compute the topological open string spectrum in the
presence of these branes. For a seven-brane, the spectrum has a
simple geometric interpretation in terms of the Dolbeault
cohomology of the $G_2$ manifold. To define the Dolbeault
cohomology, we need to use the fact that $G_2\subset SO(7)$ acts
naturally on differential forms, and we can decompose them into
$G_2$ representations. In this paper, we will use the notation
$\pi^p_{\bf n}$ to denote the projection of the space of $p$-forms
$\Lambda^p$ onto the irreducible representation ${\bf n}$ of
$G_2$. The Dolbeault complex is then
\begin{equation}
0 \longrightarrow \Lambda^0 \stackrel{d}{\longrightarrow}
\Lambda^1 \stackrel{\pi^2_{\bf 7} d}{\longrightarrow} \pi^2_{\bf
7}(\Lambda^2) \stackrel{\pi^3_{\bf 1} d}{\longrightarrow}
\pi^3_{\bf 1}(\Lambda^3) \longrightarrow 0 \; .
\end{equation}
The topological open string BRST cohomology is the cohomology of
this complex and yields states at ghost numbers $0,1,2,3$. For
zero-, three- and four-branes the cohomology is obtained by
reducing the above complex to the brane in question.

In section~4 we will verify explicitly that the BRST cohomology in
ghost number one contains the space of (generalized) flat
connections on the brane, but also contains the infinitesimal
moduli of the topological brane. In particular, we will see that
the topological open string reproduces precisely the results in
the mathematics literature \cite{McLean:1998} regarding
deformations of calibrated cycles in manifolds of $G_2$ holonomy.

We briefly discuss scattering amplitudes in section~5 and use them
to construct the open topological string field theory following
methods discussed in \cite{Witten:1992fb} in section~6. The final
answer for the open topological string field theory turns out to
be very simple. For seven-branes we obtain the associative
Chern-Simons (CS) action
\begin{equation}
S =\int_Y \ast\phi\wedge CS_3(A) \; ,  \label{act}
\end{equation}
with $CS_3 (A)$ the standard Chern-Simons three-form and
$\ast\phi$ the harmonic four-form on the $G_2$ manifold $Y$. For
the other branes we obtain the dimensional reduction of this
action to the appropriate brane. The action (\ref{act}) was first
considered in \cite{Donaldson:1996kp,Baulieu:1997nj}, and it is
gratifying to have a direct derivation of this action from string
theory. We will also discuss the dimensional reduction of this
theory on $CY_3 \times S^1$, which leads to various real versions
of the open A- and B-model, depending on the brane one is looking
at. The situation is very similar to the closed topological $G_2$
string, which also reduced to a combination of real versions of
the A- and B-models. It is presently unclear to us whether we
should interpret this as meaning that the partition functions of
the open and closed topological $G_2$ strings should not be
interpreted as wave functions, as opposed to the partition
functions of the open and closed A- and B-models, which are most
naturally viewed as wave functions.

The last subject we discuss in section~6 is the emergence of
worldsheet instanton contributions of the topological string
theory on Calabi-Yau manifolds from the topological $G_2$ string
on $CY_3 \times S^1$. Though our analysis is not yet conclusive,
it appears that these worldsheet instanton effects arise from
wrapped branes in the $G_2$ theory and not directly from
worldsheet instantons.

Finally, in section 7 we make a preliminary investigation of the gauge-fixing
and quantization of (\ref{act}) and its reductions to four- and
three-dimensional branes. As was the case in the open string field theory for
A-branes, the gauge-fixed actions look very similar to the action (\ref{act})
once we replace the ghost number one field $A$ by a field of arbitrary ghost
number. We also study the one-loop partition functions of the various open
string field theories, and find that they tend to have the effect of shifting
the tree-level theories in a rather simple way. This is similar to the one-loop
shift $k \rightarrow k+h(G)$ of the level $k$ in ordinary Chern-Simons theory,
with $h(G)$ the dual Coxeter number of the gauge group $G$. In particular, we
find that ${*\phi}$ in (\ref{act}) is shifted by a four-form proportional to the
first Pontrjagin class of the manifold $Y$.  We have not yet attempted to
determine whether (\ref{act}) is renormalizable and well-defined as a quantum
theory (which, by naive power-counting, it is not) but we expect that it should
be as it is equivalent to a string theory (a similar issue occurs for
holomorphic Chern-Simons in the B-model).

We conclude with a list of open problems and have collected
various technical results in the appendices.

We will adhere to the following conventions: $M$ will refer to a
calibrated submanifold of dimension 3 or 4 (i.e. calibrated by
$\phi$ or $*\phi$, respectively); these are known, respectively,
as associative and coassociative submanifolds.  The ambient $G_2$
manifold will be denoted $Y$.

\section{A brief review of the closed topological $G_2$ string}

Let us briefly review the definition of the topological $G_2$
string found in \cite{deBoer:2005pt}.  We will cover only
essential points.  For further details we refer the reader to
\cite{deBoer:2005pt}.

\subsection{Sigma model for the $G_2$ string}

The topological $G_2$ string constructs a topological string
theory with target space a seven-dimensional $G_2$-holonomy
manifold $Y$. This topological string theory is defined in terms
of a topological twist of the relevant sigma-model. In order to
have $\CN=1$ target space supersymmetry, one starts with an $\CN =
(1,1)$ sigma model on a $G_2$ holonomy manifold. The special
holonomy of the target space implies an extended supersymmetry
algebra for the worldsheet sigma-model \cite{Howe:1991ic}. That
is, additional conserved supercurrents are generated by pulling
back the covariantly constant 3-form $\phi$ and its hodge dual
$*\phi$ to the worldsheet as
$$  \phi_{\mu\nu\rh}(\sX) D\sX^\mu D \sX^\nu D \sX^\rh \; , $$
where $\sX$ is a worldsheet chiral superfield, whose bosonic
component corresponds to the world-sheet embedding map. From the
classical theory it is then postulated that the extended symmetry
algebra survives quantization, and is present in the quantum
theory. This postulate is also based on analyzing all possible
quantum extensions of the symmetry algebra compatible with
spacetime supersymmetry and $G_2$ holonomy.

A crucial property of the extended symmetry algebra is that it contains an $\CN
=1$ SCFT sub-algebra, which has the correct central charge of c=7/10 to
correspond to the tri-critical Ising unitary minimal model.  Unitary minimal
models have central charges in the series $c = 1 - \frac{6}{p(p+1)}$ (for
$p$ an integer) so the tri-critical Ising model has $p=4$.

The conformal primaries for such models are labelled by two integer Kac labels,
$n'$ and $n$, as $\phi_{(n',n)}$ where $1 \leq n' \leq p$ and $1 \leq n < p$.
The Kac labels determine the conformal weight of the state as $h_{n',n} =
\frac{[pn' - (p+1)n]^2 -1}{4p(p+1)}$.  The Kac table for this minimal model is
reproduced in \cite[Table 1]{deBoer:2005pt}.  Note that primaries with label
$(n', n)$ and $(p+1 -n', p-n)$ are equivalent.  This model has six conformal
primaries with weights $h_I =0,1/10, 6/10, 3/2$ (for the NS states) and
$h_I=7/16, 3/80$ (for the R states). 

The conformal block structure of the weight 1/10, $\phi_{(2,1)}$, and of the
weight 7/16 primary, $\phi_{(1,2)}$, is particularly simple,

\begin{eqnarray}
\phi_{(2,1)} \times \phi_{(n',n)} &=& \phi_{(n'-1, n)} +
\phi_{(n'+1, n)} \; ,
\nonumber \\
\phi_{(1,2)} \times \phi_{(n',n)} &=& \phi_{(n', n-1)} +
\phi_{(n', n+1)} \; ,\nonumber
\end{eqnarray}

\noindent where $\phi_{(n',n)}$ is any primary. This conformal block decomposition is
schematically denoted as
\begin{eqnarray}
\Phi_{(2,1)} &=& \Phi_{(2,1)}^{\downarrow} \oplus
\Phi_{(2,1)}^{\uparrow} \; ,
\nonumber \\
\Phi_{(1,2)} &=& \Phi_{(1,2)}^{-} \oplus \Phi_{(1,2)}^{+} \; .
\end{eqnarray}

\noindent The conformal primaries of the full sigma-model are
labelled by their tri-critical Ising model highest weight, $h_I$,
and the highest weight corresponding to the rest of the algebra,
$h_r$, as $|h_I, h_r \rangle$.  This is possible because the stress
tensors, $T_I$, of the tricritical sub-algebra and of the \lq rest'
of the algebra, $T_r = T - T_I$ (where $T$ is the stress tensor of
the full algebra), satisfy $T_I \cdot T_r \sim 0$.

\subsection{The $G_2$ twist}

The standard $\CN = (2,2)$ sigma-models can be twisted by making
use of the U(1) R-symmetry of their algebra. Using the U(1)
symmetry, the twisting can be regarded as changing the worldsheet
sigma-model with a Calabi-Yau target space by the addition of the
following term:
\begin{equation}\label{eqn_twist_insert}
  \pm \frac{\om}{2} \psib \psi \; ,
\end{equation}
with $\omega$ the spin connection on the world-sheet. This
effectively changes the charge of the fermions under worldsheet
gravity to be integral, resulting in the topological A/B-model
depending on the relative sign of the twist in the left and right
sector of the theory (for fermions with holomorphic or
anti-holomorphic target space indices). Here $\psib$ and $\psi$ can
be either left- or right-moving worldsheet fermions and $\om$ is
the spin-connection on the worldsheet.  In the topological theory,
before coupling to gravity, there are no ghosts or anti-ghosts so
these are the only spinors/fermions in the system.

This twist has been re-interpreted \cite{Bershadsky:1993cx,
Witten:1991zz} as follows.  First think of the exponentiation of
(\ref{eqn_twist_insert}) as an insertion in the path integral
rather than a modification of the action.  By bosonising the
world-sheet fermions we can write $\psib \psi = \prt H$ for a free
boson field so the above becomes

\begin{equation}
  \int \frac{\om}{2} \prt H =  - \int H \frac{\prt \om}{2} = \int H
  R\; ,
\end{equation}

\noindent where $R$ is the curvature of the world-sheet. We can
always choose a gauge for the metric such that $R$ will only have
support on a number of points given by the Euler number of the
worldsheet.

For closed strings on a sphere the Euler class has support on  two points which
can be chosen to be at $0$ and $\infty$ (in the CFT defined on the sphere) so
the correlation functions in the topological theory can be calculated in terms
of the original CFT using the following dictionary:

\begin{equation}
  \braket{\dots}_{twisted} = \braket{e^{H(\infty)}\dots
  e^{H(0)}}_{untwisted}\; .
\end{equation}

\noindent The \lq untwisted' theory should not be confused with
the physical theory, because it does not include integration over
world-sheet metrics and hence has no ghost or superghost system
and also it is still not at the critical dimension.  The equation
above simply relates the original untwisted $\CN = 2$ sigma-model
theory to the twisted one.

In \cite{deBoer:2005pt} a related prescription is given to define
the twisted \lq topological' sigma-model on a 7-dimensional target
space with $G_2$ holonomy. Here the role of the U(1) R-symmetry is
played by the tri-critical Ising model sub-algebra. However, a
difference is that the topological $G_2$ sigma-model is formulated
in terms of conformal blocks rather than in terms of local
operators. In particular the operator $H$ in the above is replaced
by the conformal block $\Phi^{+}_{(1,2)}$.

The main point of the topological twisting is to redefine the theory in such a
way that it contains a scalar BRST operator. In the $G_2$ sigma model, the BRST
operator is related to the conformal block of the weight 3/2 current $G(z)$ of
the super stress-energy tensor\footnote{The super stress-energy tensor is given
as  ${\mathbf T}(z, \theta) = G(z) + \theta T(z)$. The current G(z) can be
further decomposed as G(z) = $\Phi_{(2,1)} \otimes \Psi_{{14 \over 10}}$, in
terms of the tri-critical Ising-model part and the rest of the algebra,
respectively. Since its tri-critical Ising model part contains only the primary
$\Phi_{(2,1)}$, it can be decomposed into conformal blocks accordingly.}, $$ Q
= G_{-{1 \over 2}}^{\downarrow} \; .$$ It should be pointed out that in
\cite{deBoer:2005pt} it was not possible to explicitly construct the twisted
stress tensor, and although there is circumstantial evidence that the
topological theory does exist beyond tree level this statement remains
conjectural.

\subsection{The $G_2$ string Hilbert space}\label{subsec_g2_hilb}

In a general CFT the set of states can be generated by acting with
primary operators and their decendants on the vacuum state,
resulting in an infinite dimensional Fock space.  In string sigma
models this Fock space contains unphysical states, and so the
physical Hilbert space is given by the cohomology of the BRST
operator on this physical Hilbert space which is still generally
infinite-dimensional.

In the topological A- and B-models a localization argument
\cite{Witten:1991zz} implies that only BRST fixed-points
contribute to the path integral and these correspond to
holomorphic and constant maps, respectively.  Thus the set of
field configurations that when quantized, generate states in the
Hilbert space is restricted to this subclass of all field
configurations and so the Fock space is much smaller.  Upon
passing to BRST cohomology this space actually becomes
finite-dimensional.

In the $G_2$ string the localization argument cannot be made
rigorous, because the action of the BRST operator on the
worldsheet fields is inherently quantum, and so is not well
defined on the classical fields.  Neglecting this issue and
proceeding naively, however, one can construct a localization
argument for $G_2$ strings that suggests that the path integral
localizes on the space of constant maps \cite{deBoer:2005pt}.
Thus we will take our initial Hilbert space to consist of states
generated by constant modes $X^\mu_0$ and $\psi^\mu_0$ on the
world-sheet (in the NS-sector there is no constant fermionic mode
but the lowest energy mode $\psi^\mu_{-\frac{1}{2}}$ is used
instead).  These correspond to solutions of worldsheet equations
of motion with minimal action which dominate the path integral in
the large volume limit.

In \cite{Witten:1991zz} the fact that the path integral can be
evaluated by restricting to the space of BRST fixed points is
related to another feature of the A/B-models: namely the
coupling-invariance (modulo topological terms) of the worldsheet
path integral.  Variations of the path integral with respect to
the inverse string coupling constant $t \propto (\al')^{-1}$ are
$Q$-exact, so one may freely take the weak coupling limit $t \ra
\infty$ in which the classical configurations dominate.  This
limit is equivalent to rescaling the target space metric, and so
we will refer to it as the large volume limit.

Accordingly, all the calculations in the A- and B- model can be
performed in the limit where the Calabi-Yau space has a large
volume relative to the string scale, and the worldsheet theory can
be approximated by a free theory.  The $G_2$ string also has the
characteristics of a topological theory, such as correlators being
independent of the operator's positions, and the fact that the
BRST cohomology corresponds to chiral primaries.  On the other
hand since the theory is defined in terms of the conformal blocks,
it is difficult to explicitly check the coupling constant independence.
Based on the topological arguments, and on the postulate of the
quantum symmetry algebra, in this paper we will assume the
coupling constant independence and the validity of localization
arguments.  Even if these arguments should fail for subtle
reasons, the results of this paper are always valid in the large
volume limit.

\subsection{The $G_2$ string and geometry}\label{subsec_g2_geom}

As in the topological A- and B-model, for the topological $G_2$
string there is a one-to-one correspondence between local
operators of the form $O_{\omega_p}= \omega_{i_1 \ldots i_p}
\psi^{i_1} \ldots \psi^{i_p}$ and target space $p$-forms $\omega_p
= \omega_{i_1 \ldots i_p} dx^{i_1} \wedge \ldots \wedge dx^{i_p}$.
In \cite{deBoer:2005pt} it is found that the BRST cohomology of
the left (right) sector alone maps to a certain refinement of the
de Rham  cohomology described by the \lq $G_2$ Dolbeault' complex

\begin{equation}\label{eqn_g2_complex}
0 \rightarrow \Lambda^{0}_{\bf 1} \xrightarrow{\check{D}}
\Lambda^1_{\bf 7} \xrightarrow{\check{D}} \Lambda^2_{\bf 7}
\xrightarrow{\check{D}} \Lambda^3_{\bf 1} \rightarrow 0\; .
\end{equation}

The notation is that $\Lambda^p_{\bf n}$ denotes differential forms of degree
$p$, transforming in the irreducible representation ${\bf n}$ of $G_2$. The
operator $\check{D}$ acts as the exterior derivative on 0-forms, and as
\begin{eqnarray} {\check D} (\alpha) &=& \pi^2_{\bf 7} (d \alpha) \quad {\rm if}
  \quad \alpha \in \Lambda^1 \; , \nonumber \\ {\check D} (\beta) &=& \pi^3_{\bf
  1} (d \beta)  \quad {\rm if} \quad \beta \in \Lambda^2  \; , \nonumber
\end{eqnarray} where $\pi^2_{\bf 7}$ and $\pi^3_{\bf 1}$ are projectors onto the
relevant representations. The explicit expressions for the projectors and the
standard decomposition of the de Rham cohomology are included in appendix
\ref{app_conventions}. Thus, the BRST operator $G_{-1/2}^{\downarrow}$ maps to
the differential operator of the complex $\check{D}$.   In the closed theory,
combining the left- and right-movers, one obtains the full cohomology of the
target manifold, accounting for all geometric moduli: metric deformations, the
$B$-field moduli, and rescaling of the associative 3-form $\phi$. The relevant
cohomology for the open string states will be worked out in the following
sections.

\section{Open string cohomology}

We will now consider the $Q$ cohomology of the open string states.
Later, we will interpret part of this cohomology in terms of
geometric and non-geometric (gauge field) moduli on calibrated 3-
and 4-cycles.  

In \cite{deBoer:2005pt} states in the $G_2$ CFT were shown to satisfy a certain
non-linear bound in terms of $h_I$ and $h_r$ and states saturating this bound
are argued to fall into shorter, BPS, representation of the non-linear $G_2$
operator algebra. Such states are referred to as chiral primaries.  Analogous to
the $\CN = 2$ case, it is the physics of these primaries that the twist is
intended to capture and thus they are the states that occur in the BRST
cohomology.  The chiral primaries in the NS sector have $h = 0, 1/2, 1, 3/2$ and
$h_I = 0, 1/10, 6/10, 3/2$ and they are the image of the RR ground states under
spectral flow.

Recall that we are working in the zero mode approximation (corresponding to the
large volume limit, $t\ra \infty$,  where oscillator modes can be neglected) and
in this limit a general state is of the form $A_{\mu_1 \dots \mu_n} (X_0)
\psi_0^{\mu_1} \dots \psi_0^{\mu_n}$.  On such states $L_0$ acts as $t \Box +
\frac{n}{2}$ so states with $h = 0, 1/2, 1, 3/2$ correspond to 0, 1, 2, and 3
forms ($f(X_0)$, $A_\mu(X_0) \psi^\mu_0$, \dots).  As argued in
\cite{deBoer:2005pt} we can thus consider $Q$-cohomology on the space of 0, 1,
2, and 3 forms restricted to those that have $h_I = 0, 1/10, 6/10, 3/2$,
respectively.

In general we are interested in harmonic representatives of the $Q$ cohomology
so we will look for operators (corresponding to states) that are both $Q$- and
$\Qd$-closed. The results we obtain are essentially the same as those for one
side of the closed worldsheet theory \cite{deBoer:2005pt}.

\subsection{Degree one}\label{subsec_deg_one}

We will start by looking at the $h = 1/2$ state, because it is the only one that
will generate a marginal deformation of the theory.  A general state with $h =
1/2$ is of the form $A_\mu(X)\psi^\mu$ so long as \footnote{Although we will
sometimes use the full fields $X$ and $\psi$ in the CFT and also consider OPE's
which generate deriviatives of these fields the reader should recall that we are
always working in the large volume limit where these reduce to $X_0$ and
$\psi_0$.  }

\begin{equation}
[L_0, A_\mu(X)] = t\Box A_\mu(X) = 0
\end{equation}
It also satisfies
$$[L_0^I, A_\mu(X)\psi^\mu ] = \frac{1}{10}
A_\mu(X)\psi^\mu \; ,$$ so it is a chiral primary (i.e. it
saturates the chiral bound).  Because it is a chiral primary, it
has to be $Q$-closed \cite{deBoer:2005pt}.  Rather than proceed along these
lines, however, we will consider the $Q$-cohomology directly from the definition
of $Q$.

Let us determine the $Q$-cohomology of 1-forms $\CA = A_\mu (X)\psi^\mu$.  We
first calculate $\{\Gns, A_\mu(X)\psi^\mu\}$ in the CFT on the complex plane
with $z$ complex \lq bulk' coordinates and $y$ \lq boundary' coordinates on the
real line

\begin{equation}
  \begin{split}
\{G_{-1/2}, A_\mu(X) \psi^\mu \} &= \oint dz \; G(z)  \cdot A_\mu(X) \psi^\mu(y) \; , \\
G(z)  \cdot A_\mu(X) \psi^\mu(y) &= g_{\rho\sg}(X) \psi^\rho \prt X^\sg(z) \cdot A_\mu(X)
\psi^\mu(y) \\
     &\sim \prt(\ln |z - y|^2 + \ln |\ovr{z} - y|^2)
     \nabla_\rho A_\mu \psi^\rho(z) \psi^\mu(y) \\
     &+ \frac{1}{z-y} \prt X^\mu(z) A_\mu(X(y)) \; .
\end{split}
\end{equation}

\noindent This gives\footnote{We have not been careful about the relative normalizations of the
bosonic and fermionic bulk-boundary OPE's, but this is not relevant as in all
computations of this type that occur below, we will only end up keeping one of
the terms.}

\begin{equation}
 \{G_{-1/2}, A_\mu(X) \psi^\mu \}=  A_\mu \prt X^\mu(y) + \frac{1}{2} \prt_{[\mu} A_{\nu]} \psi^\mu \psi^\nu \; .
\end{equation}

To compute the action of $Q$ we now project onto the $\dnar$ part, which
includes only the part with tri-critical Ising weight $6/10$.  The term $A_\mu
\prt X^\mu$  vanishes in the zero mode limit so we only need to consider the
second term.  The condition that this term has $h_I = \frac{6}{10}$ is
\cite{deBoer:2005pt}

\begin{equation}
(\pi_{\bf 14}^2)^{\rho\sg}_{\phnt{\rho\sg}\mu\nu} \prt_{[\rho}
A_{\sg]} = 0 \; ,
\end{equation}

\noindent where $\pi_{\bf 14}^2$ is the projector onto the 2-form
subspace $\Lm^2_{\bf 14} \subset \Lm^2$, in the {\bf 14}
representation of $G_2$.

This result implies that the $\frac{6}{10}$ part of $\prt_{[\rho}
A_{\sg]}$ (or any 2-form) is in $\Lm_{\bf 7}^2$,
so on a 1-form we can define $Q$ as

\begin{equation}\label{eqn_Q_closure}
  \{Q, A_\mu\psi^\mu\} = (\pi_{\bf 7}^2) \{\Gns, A_\mu\psi^\mu\} = 6
  \phi_{\mu\nu}^{\phnt{\mu\nu}\gm}\phi_{\gm}^{\phnt{\gm}\rho\sg}  \prt_{[\rho}
  A_{\sg]} dx^\mu \wedge dx^\nu   = \check{D} A =
  0\; ,
\end{equation}

\noindent where we have used

\begin{equation}\label{eqn_7_proj}
  (\pi_{\bf 7}^2)^{\rho\sg}_{\phnt{\rho\sg}\mu\nu} = 4 (*\phi)^{\rho\sg}_{\phnt{\rho\sg}\mu\nu} + \frac{1}{6}
  (\dl^\rho_\mu \dl^\sg_\nu - \dl^\sg_\mu \dl^\rho_\nu) = 6 \phi_{\mu\nu}^{\phnt{\mu\nu}\gm}\phi_{\gm}
^{\phnt{\gm}\rho\sg} \; .
\end{equation}

\noindent Note that $Q$ acting on 1-forms has reduced essentially to
$\check{D}$; the same will occur for forms of other degrees.

Let us now consider $Q$-coclosure.  The inner product of
states

$$\braket{A_{[\mu\nu]}\psi^\mu \psi^\nu|B_{[\al\bt]} \psi^\al
\psi^\bt} \; ,$$

\noindent becomes the inner product of forms $\int_Y (*A \wedge
B)$, so $Q^\dag$ acting on $A$ is given by \newline $\braket{Q
\cdot f(X) |A_\mu(X) \psi^\mu} = \braket{f(X)| Q^\dag \cdot A_\mu
\psi^\mu}$, which can be determined as

\begin{equation}
 % \begin{split}
 %&
 \braket{Q \cdot f(X) |A_\mu(X) \psi^\mu} = \int \sqrt{g} \prt_\mu f(X)
  A^\mu(X) \\
  %&
 %= - \int f(X) \prt_\mu(\sqrt{g}A^\mu(X))
 = - \int \sqrt{g} f(X)
  \nabla_\mu A^\mu(X) \; .
 % \end{split}
\end{equation}

\noindent So if $A_\mu$ is also required to satisfy

\begin{equation}\label{eqn_Qdag_closure}
Q^\dag \cdot A_\mu(X) \psi^\mu = - \nabla_\mu A^\mu(X) = 0 \; ,
\end{equation}

\noindent then it is $Q$- and $Q^\dag$-closed and hence a harmonic
represenative of $Q$-cohomology.

\subsection{Degree zero}

The cohomology in degree zero is rather trivial.  Given a degree
zero mode $f(X)$ we have $\{Q, f(X)\} = \prt_\mu f(x) \psi^\mu$.
This follows from $Q = \Gdn = \Gns$,  because the projection onto
the $\downarrow$ component is trivial since all operators of the
form $A_\mu(X) \psi^\mu$ automatically have $L_0^I$ weight
$\frac{1}{10}$.  So $Q$-closure implies

\begin{equation}
\prt_\mu f(X) = 0 \; .
\end{equation}

\noindent The $\Qd$-closure here is vacuous since there are no lower degree fields.

\subsection{Degree two}

In degree two we start with a two form $\om_{\rho \sg} \psi^\rho
\psi^\sg$ which should have $L_0^I$ weight $\frac{6}{10}$, so it
should satisfy $\pi^2_{\bf 7}(\om) = \om$.  The need to restrict
$\om \in \Lm^2_{\bf 7}$ comes from the way $Q$ is defined in
\cite{deBoer:2005pt}.  We must once more calculate the action of
$\Gns$,
and then project it onto the $\downarrow$ part

\begin{align}
  \{\Gns, \om\} &= \oint dz \; g_{\mu\nu}\psi^\mu \prt X^\nu(z) \cdot \om_{\rho
  \sg} \psi^\rho \psi^\sg \nonumber \\
  &= \oint dz \; \frac{1}{z} g_{\mu\nu}\prt^\nu \om_{\rho \sg} \psi^\mu \psi^\rho \psi^\sg
  + \frac{1}{z} g_{\mu\nu} \prt X^\nu \om_{\rho \sg} g^{\mu\rho} \psi^\sg
  - \frac{1}{z} g_{\mu\nu} \prt X^\nu \om_{\rho \sg} g^{\mu\sg} \psi^\rho \nonumber \\
 &= \prt_\mu \om_{\rho \sg} \psi^\mu \psi^\rho \psi^\sg + 2 \om_{\rho \sg} \prt
 X^\rho  \psi^\sg \; .
\end{align}

\noindent  Once more we can drop the second term in the large volume limit in
which we are working.  We use the result in \cite{deBoer:2005pt} that the
projector onto the $L_0^I$ weight $\frac{3}{2}$ corresponds to the projector
onto $\Lm^3_{\bf 1}$, and is given by contracting with the associative 3-form
$\phi$. So for $\Om \in \Lm^3_{\bf 1}$

\begin{equation}
  \phi^{\al\bt\gm}\Om_{\al\bt\gm} \phi_{\mu\nu\rho} = 7 \Om_{\mu\nu\rho} \; .
\end{equation}

\noindent In particular, we can project onto the $\frac{3}{2}$ part of
$\{\Gns, \om\} = \prt_\mu \om_{\rho \sg}$  using
$\phi^{\al\bt\gm}$, so $Q$-closure implies

\begin{equation}\label{eqn_Q2_closure}
  \phi^{\al\bt\gm} \prt_{[\al} \om_{\bt \gm]} = 0 \; .
\end{equation}

\noindent Note that this once again can be written as $\check{D}\om = 0$.

We will now derive the $\Qd$-closure condition.  This is done in
exactly the same way as was done for the degree one components

\begin{equation}
  \begin{split}
  \braket{\om |Q \cdot A_\mu(X) \psi^\mu}
  & = \int \sqrt{g} \om^{\mu\nu} (\pi_{\bf 7}^2)^{\al\bt}_{\phnt{\al\bt}\mu\nu} \prt_{\al} A_{\bt}
  = - \int  \sqrt{g} A_{\bt}  \bigl( (\pi_{\bf 7}^2)^{\al\bt}_{\phnt{\al\bt}\mu\nu}
  \nabla_\al \om^{\mu\nu} \bigr) \; ,
  \end{split}
\end{equation}

\noindent so

\begin{equation}
\Qd \cdot \om = -(\pi_{\bf 7}^2)^{\mu\nu}_{\phnt{\mu\nu}\al\bt}
\nabla^\al \om_{\mu\nu}
 dx^\beta = -6\phi^{\mu\nu}_{\phnt{\mu\nu}\gm} \phi^{\gm}_{\phnt{\gm}\al\bt}  \nabla^\al
 \om_{\mu\nu} dx^\beta = -\nabla^\al \om_{\al\bt} dx^\beta = 0 \; .
\end{equation}

\noindent Here we have used $\pi^2_{\bf 7}(\om) = \om$.

\subsection{Degree three}

A 3-form $\Om_{\mu\nu\rho}\psi^\mu \psi^\nu \psi^\rho$ is first projected onto
its $\Lm^3_{\bf 1}$ component by $Q$, so we take $\pi^3_{\bf 1}(\Om) = \Om$,
which means that $\Om$ is a function times $\phi$.  From the definition of $Q$
it is evident that it acts trivially on $\Om$ since there is no higher $L_0^I$
eigenstate in the NS sector for $Q$ to project onto.  This implies $Q = 0$ on
three forms which matches (\ref{eqn_g2_complex}).  Thus we see that the action
of $Q$ on states in the zero mode approximation maps into the complex
(\ref{eqn_g2_complex}) as anticipated in Section \ref{subsec_g2_geom}.

The $Q$-coclosure of $\Om$ is derived similarly to the 1- and
2-form case and gives

\begin{equation}
  Q^\dag \cdot \Om = \nabla^\mu \Om_{\mu\nu \rho} dx^\nu \wedge dx^\rho = 0 \; .
\end{equation}

\subsection{Harmonic constraints}

In the previous subsections we considered the conditions for $Q$-
and $Q^\dag$-closure on the states in the $G_2$ CFT.  These
conditions are all linear in derivatives but they must be enforced
simultaneously to generate unique representatives of
$Q$-cohomology.  As $Q$ corresponds to the operator ${\check D}$
discussed in section \ref{subsec_g2_geom}, it generates the
Dolbeault complex (\ref{eqn_g2_complex}) which is known to be
elliptic \cite{ReyesCarrion:1998si, Donaldson:1996kp} and so can
be studied using Hodge theory. This implies that physical states
in the theory correspond to the kernel of the Laplacian operator
$\{Q, Q^\dag\}$, so one can equivalently consider this single
non-linear condition instead of the two seperate linear conditions
imposed by $Q$ and $Q^\dag$.

These $Q$-harmonic conditions (derived from the actions of $Q$ and
$Q^\dag$) are

\begin{align}
  \{Q, Q^\dag\} \cdot f &= \nabla_\mu \prt^\mu f =0 \; , \nonumber \\
  \{Q, Q^\dag\} \cdot A_\nu \psi^\nu &= \bigl(\nabla_\nu \nabla_\mu A^\mu  +
  (\pi^2_{\bf 7})_{\nu}^{\phnt{\nu}\gm\mu\sg} \nabla_\gm \nabla_\mu A_\sg
  \bigr) \psi^\nu =0 \; , \label{eqn_deg1_harm}  \nonumber  \\
  \{Q, Q^\dag\} \cdot \om_{\mu \nu} \psi^\mu \psi^\nu &=
  \bigl( (\pi^2_{\bf 7})_{\mu\nu}^{\phnt{\mu\nu}\al\bt}
  \nabla_\al \nabla^\gm \om_{\bt\gm} +
  (\pi^3_{\bf 1})_{\mu\nu\rh}^{\phnt{\mu\nu\gm}\al\bt\gm} \nabla^\rh \nabla_\al
  \om_{\bt\gm} \bigr) \psi^\mu \psi^\nu =0 \; .
\end{align}

\noindent We have used $\pi^2_7(\om) = \om$ to simplify the last
expression above.

\section{Open string moduli}\label{subsec_moduli}

In a general topological theory one can use elements of degree one
cohomology to deform the theory using descendant operators.  If
$\CO$ is a degree one operator, in the A/B-model this means that
it has ghost number one, whereas in the $G_2$ string this means
that it corresponds to one \lq +' conformal block. Then one can
deform the action by adding a term $\int_{\prt \Sg} \{\Gup,
\CO\}$, which is $Q = \Gdn$ closed and of degree 0.  Thus the
elements of $H_Q^1$ cohomology should correspond to possible
deformations of the theory or tangent vectors to the moduli space
of open topological $G_2$  strings.

Since open strings correspond to supersymmetric\footnote{In the
sense of preserving the extended worldsheet superalgebra.} branes,
the full moduli space  should include both the moduli space of the
field theory on the brane as well as the geometric moduli of the
branes.  For $G_2$ manifolds the latter are simply the moduli of
associative and coassociative 3- and 4-cycles, respectively, which
have been studied in \cite{McLean:1998}.  Below we will show that
the operators $\CO$
 corresponding to normal modes do satisfy the correct constraints to be
deformations of the relevant calibrated submanifolds.  Since a
priori it is not known what the field theory on these branes will
be, in the topological case we will study the constraints on the
tangential modes (which in physical strings would correspond to
gauge fields on the brane), and attempt to interpret these as
infinitesimal deformations in the moduli space of some gauge
theory on the brane.

\subsection{Calibrated geometry}

In order to preserve the extended symmetry algebra (such as $\CN =
2$ or $G_2$) of the worldsheet SCFT in the presence of a boundary,
certain constraints must be imposed on the worldsheet currents.
These have been studied in \cite{Ooguri:1996ck}
\cite{Becker:1996ay}, and more extensively in
\cite{Albertsson:2001dv} \cite{Albertsson:2002qc}
\cite{Howe:2005je}.  One imposes the boundary condition on the
left- and right-moving components of the worldsheet fermions,
$\psi_L^\mu = R^\mu_\nu(X) \psi^\nu_R$, and then conservation of
the worldsheet currents in the presence of the boundary implies
that, on the subspace $M$ where open strings can end,

\begin{equation}
  \begin{split}
  \phi_{\mu\nu\sg} &= \eta_\phi R^\al_\mu R^\bt_\nu R^\gm_\sg \phi_{\al\bt\gm} \; , \\
  (*\phi)_{\mu\nu\sg\lm} &= \eta_\phi R^\al_\mu R^\bt_\nu R^\gm_\sg R^\rho_\lm
  (*\phi)_{\al\bt\gm\lm}\det(R) \\
  &=  R^\al_\mu R^\bt_\nu R^\gm_\sg R^\rho_\lm
  (*\phi)_{\al\bt\gm\lm} \; .
  \end{split}
\end{equation}

\noindent Note that $R^\al_\mu(X)$ (for any $X \in M$) is
generally a position-dependent invertible matrix, but locally it
can be diagonalized with eigenvalues $+1$ in Neumann directions
and $-1$ in Dirichlet directions. $\eta_\phi = \pm 1$ gives two
different possible boundary conditions with the choice of
$\eta_\phi = 1$ corresponding to open strings ending on a
calibrated 3-cycle, while $\eta_\phi = -1$ corresponds to strings
on a calibrated 4-cycle \cite{Becker:1996ay}. Calibrated
submanifolds, first studied in \cite{Harvey:1982xk}, are
characterized by the property that their volume form induced by
the metric in the ambient space is the pull-back of particular
global forms, in this case
$\phi$ (for associative 3-cycles) or $*\phi$ (for coassociative 4-cycles). This
implies the volume of the calibrated submanifold is minimal in its homology class. \\
\\
\noindent{\bf Remark.}  There are several subtleties regarding
boundary conditions in topological sigma-models that deserve to be
mentioned. Below, we will advocate the perspective that any
boundary condition preserving the extended algebra\footnote{To be
precise the boundary conditions preserve some linear combination
of the extended algebra in the left/right sector of the
worldsheet. So a brane may reduce an $\CN = (2, 2)$ theory to an
$\CN = 2$ theory.} should also be a boundary condition of the
topological theory, because the presence of an extended algebra
allows one to define a twisted theory.  In the A- and B-model,
however, although both the A- and B-brane boundary conditions
preserve the $\CN = 2$ algebra, each is compatible with only one
of the twists, so a given topological twist is not necessarily
compatible with an arbitrary algebra-preserving boundary
condition. Moreover, a given topological twist might only depend
on the existence of a subalgebra of the full extended algebra, so
may be possible even with boundary conditions that do not preserve
the full extended algebra. A concrete example of this is the
Lagrangian boundary condition for the A-model branes proposed by
Witten \cite{Witten:1992fb}. This condition is considerably less
restrictive that the special Lagrangian condition required to
preserve the full $\CN = 2$ algebra in the physical string
\cite{Ooguri:1996ck} and reflects the fact that the A-model is
well-defined for any K\"{a}hler manifold and does not require a
strict Calabi-Yau target space.  While similar subtleties might,
in principle, exist for the $G_2$ twist they are concealed by the
fact that the twist does not have a classical realization that we
know of. So we will tentatively assume the correct boundary
conditions are those that preserve the full $G_2$ algebra on one
half of the worldsheet theory.

\subsection{Normal modes}\label{sec_normal_modes}

Let us now consider the cohomology of open strings ending on a
D-brane which wraps either an associative 3-cycle or a
co-associative 4-cycle.  We adopt the convention that $I, J, K,
\dots$ are indices normal to the brane while $a, b, c, \dots$ are
tangential, and Greek letters run over all indices.  \noindent The
state $A_\mu \psi^\mu$ decomposes into normal and tangential modes
which will be denoted $\tht_I \psi^I$ and $A_a \psi^a$
respectively; all momenta is tangential, denoted by $k_a$.  The
normal modes will have the form $\CA = \tht_I(X^a) \psi^I$ so
$G_{-1/2} \cdot \CA = \prt_a \tht_I(X^b) \psi^a \psi^I$.  Here
$\CA$ will denote a {\em general} operator/state in the CFT and
should not be confused with the gauge field (or operator) $A_\mu
\psi^\mu$.

\paragraph{Associative 3-cycles.}

Let us now consider the $Q$-cohomology when restricted to an
associative 3-cycle $M$. On the 3-cycle the form $\phi$ must
satisfy \cite{Howe:2005je}

\begin{equation}
  \begin{split}
  \phi_{\mu\nu\sg} &= R^\al_\mu R^\bt_\nu R^\gm_\sg \phi_{\al\bt\gm} \; . \\
  \end{split}
\end{equation}

\noindent Since $M$ is associative, $\phi$ acts as a volume form
on this cycle and, from the above, it is only non-vanishing for an
odd number of tangential indices\footnote{Here, and throughout the paper, we
will take $\ep$ to be the volume form on the (sub)manifold not merely the
antisymmetric tensor.}

\begin{equation}
  \begin{split}
  \phi_{abc} &= \ep_{abc}  \; ,\\
  \phi_{Ibc} &= 0 \; ,\\
  \phi_{IJK} &= 0 \; .
  \end{split}
\end{equation}

\noindent The $Q$-closure of normal modes is given by (\ref{eqn_Q_closure})

\begin{equation}\label{eqn_3c_norm_cond}
\phi_{bK}^{\phnt{bK}J}\phi_{J}^{\phnt{J}aI}  \nabla_{a} \tht_{I} =
0 \; ,
\end{equation}

\noindent where the index structure is enforced by the requirement that $\phi$
has an even number of normal indices.

To understand the geometric significance of equation
(\ref{eqn_3c_norm_cond}) in the abelian theory, recall that
$\tht^I$ is just a section of the normal bundle $NM$ of $M$ in
$Y$, which by the tubular neighborhood theorem can be identified
with an infinitesimal deformation of $M$.  This equation is the
linear condition on $\tht^I$ such that the exponential map
(defined by flowing along a geodesic in $Y$ defined by $\tht^I$)
$\exp_\tht(M)$ takes $M$ to a new associative submanifold $M'$.
This is just a reformulation of the condition given in
\cite{McLean:1998}.

In \cite{McLean:1998} McLean defines a functional on the space of (integrable)
normal bundle sections by

\begin{equation}
  F_{\gm}(\tht) = (*\phi(x))_{\mu\nu\rh\gm} \frac{\prt x^\mu}{\prt \sg^a}
  \frac{\prt x^\nu}{\prt \sg^b} \frac{\prt x^\rh}{\prt \sg^c}
  \epsilon^{abc} \propto (*\phi(x))_{\mu\nu\rh\gm} \frac{\prt x^\mu}{\prt \sg^a}
  \frac{\prt x^\nu}{\prt \sg^b} \frac{\prt x^\rh}{\prt \sg^c} \phi^{abc} \; .
\end{equation}

\noindent Here $x(t, \tht, \sg) = {\mbox{exp}}_\tht(\sg, t)$ is a geodesic curve
parameterized by the variable $0 < t < t_1$, which starts at a point $\sg \in M$
with $\dot{x}(\sg) = \tht$ at $t=0$, and flows after a fixed time to $x(t=t_1,
\tht, \sg) \in M'$, the new putative associative submanifold.  The functional is
just the pull-back\footnote{More precisely we are pulling back $\chi \in
\Om^3(Y, TY)$, a tangent bundle valued 3-form, defined using the $G_2$ metric
$\chi^\al_{\phnt{\al}\mu\nu\rh} = g^{\al\bt} (*\phi)_{\bt\mu\nu\rh}$.} of
$*\phi$ from $M'$ to $M$ and it should vanish if $M'$ is associative.

For $M'$ to be a associative it turns out to be sufficient to require that the
time derivative of $F$ at $t=0$ vanishes, which gives

\begin{equation}
  \dot{F}_{\gm}(\tht)|_{t=0} = (*\phi(x))_{Ibc\gm} \prt_a \tht^I \phi^{abc} =
  \phi_{I\gm}^{\phnt{I\gm}a} \prt_a \tht^I.
\end{equation}

\noindent This is equivalent to (\ref{eqn_3c_norm_cond}) since
each choice of $bK$ indices in that equation gives only one
non-vanishing term.
The space of such deformations is generally not a smooth manifold
and currently the moduli space of associative submanifolds of a
given $G_2$ manifold is not well understood (but see
\cite{akbulut-2004} for some recent work on this).

At first glance (\ref{eqn_3c_norm_cond}) looks like the linearized
equation (4.7) in \cite{Anguelova:2005cv} but the fields in that
action are actually embedding maps which are non-linear, whereas
the $\tht^I$ above are more closely related to linearized
fluctuations around fixed embedding maps
\footnote{In \cite{Anguelova:2005cv}, maps $x: \Sg_3 \ra Y$ from
an arbitrary three-manifold to a $G_2$ manifold are considered and
a functional which localizes on associative embeddings is defined.
There a reference associative embedding $x_0$ is chosen and used
to define a local coordinate splitting of $x^\mu$ into tangential
$x^a$ and normal $y^I$ parts. This is different from the present
situation where $\tht^I$ is an infinitesimal normal deformation of
an associative cycle. $\tht^I$ can be identified with a section of
the normal bundle (via the tubular neighborhood theorem) and is
essentially a linear object, whereas the $y^I$ above are a local
coordinate representation of a non-linear map. Basically
$\theta^I$ here are related to the linear variation $\delta y^I
|_{x_0}$ (evaluated at $x= x_0$) in \cite{Anguelova:2005cv}.}
.\\
\\
\noindent {\bf Remark.} The harmonic condition as follows from
(\ref{eqn_deg1_harm}) for normal modes is

\begin{equation}\label{eqn_normal_scnd}
  \begin{split}
  (\pi^2_{\bf 7} )^{a\phnt{I}bJ}_{\phnt{a}I} \nabla_a \nabla_b \tht_J  = 0 \; .
  \end{split}
\end{equation}
This also has a nice geometrical interpretation as vector fields
$\tht^I$ extremizing the action

\begin{equation}
  \int_M \braket{Q \cdot \tht , Q  \cdot\tht} \; ,
\end{equation}

\noindent on the associative 3-cycle. Theorem 5-3 in
\cite{McLean:1998} shows that the zeros of this action (which are
extrema since it is positive semi-definite) correspond to a family
of deformations through minimal submanifolds.

\paragraph{Coassociative 4-cycles.}

The consideration of the 4-cycle $M$ is similar to that of the
3-cycle, but now in the boundary condition we have $\eta_\phi =
-1$, so the non-vanishing components of $\phi$ must have an odd
number of normal indices and

\begin{equation}
  \begin{split}
  \phi_{abc} &= 0 \; . \\
  \end{split}
\end{equation}

\noindent Let us first consider the $Q$-closure of $\tht_I$

\begin{equation}\label{eqn_norm_cond}
\phi_{Ic}^{\phnt{\mu\nu}b}\phi_{b}^{\phnt{b}aJ}  \prt_{[a}
\tht_{J]}   = 0 \; .
\end{equation}

\noindent These are 24 equations depending on a choice of $I$ and $c$. Examining
the index structure, (\ref{eqn_norm_cond}) reduces to 4 independent equations

\begin{equation}\label{eqn_norm_cond2}
\phi_{b}^{\phnt{b}aJ}  \nabla_{a} \tht_{J}   = 0 \; ,
\end{equation}

\noindent where we replaced the commutator of a derivative with the covariant
derivative on $M$ in the induced metric.

Following \cite{McLean:1998}, let us observe an isomorphism
between the normal bundle $NM$ of the 4-cycle $M$, and the space
of self-dual 2-forms $\Lm^2_+(M)$ on $M$, given by

\begin{align}
    \tht^I &\ra \tht^I \phi_{Iab} \equiv \Om_{ab} \; , \\
    \Om_{ab} &\ra \phi^{Iab} \Om_{ab} = \phi^{Iab} \phi_{Jab} \tht^J =
    \frac{1}{6} \tht^I \; , \label{eqn_norm_sd}
\end{align}

\noindent where we have used the first identity in (\ref{eqn_g2_ident}).

To see that $\Om_{ab}$ is self-dual we use the second identity in
(\ref{eqn_g2_ident}) and the fact that ${*\phi}_{ab}^{\phnt{ab}cd}
\propto \ep_{ab}^{\phnt{ab}cd}$ on $M$, so that

\begin{equation}\label{eqn_selfdaul}
  (*_4 \Om )_{ab} \propto {*\phi}_{ab}^{\phnt{ab}cd} \Om_{cd} = \phi_{ab}^{\phnt{ab}cd} \tht^I
  \phi_{Icd} = \frac{1}{6} \phi_{Iab} \tht^I = \frac{1}{6}
  \Om_{ab}\; .
\end{equation}

\noindent Let us now use (\ref{eqn_norm_sd}) to see what
(\ref{eqn_norm_cond2}) implies for $\Om_{ab}$;

\begin{equation}
%  \begin{split}
  0 = \phi_{b}^{\phnt{b}aJ}  \nabla_{a} \phi_J^{\phnt{J}cd} \Om_{cd}   =
  \nabla_a \bigl(\phi_{b}^{\phnt{b}aJ}  \phi_J^{\phnt{J}cd} \Om_{cd}\bigr)
  = \nabla^a \bigl[\bigl( \frac{1}{9} \Om_{ba}+ \frac{1}{18} \Om_{ba} \bigr) \bigr]
  = \frac{1}{6} \nabla^a \Om_{ba} \; .
%  \end{split}
\end{equation}

This equation is just $d^\dag \Om = 0$, and since $\Om$ is
self-dual, it also implies $d\Om = 0$ so that $\Om$ must be
harmonic. Thus the $Q$-cohomology for the normal modes is given by
$\tht^I$ which map to harmonic self-dual 2-forms on $M$.

Since the $\Qd$-cohomology on the normal modes is trivial 
(eqn. (\ref{eqn_Qdag_closure}) is trivially true for normal directions), such
$\tht^I$ are $Q$-closed and co-closed, and hence $Q$-harmonic. Thus their
$Q$-cohomology is isomorphic to the de Rham cohomology group $H^2_+(M)$ of
harmonic self-dual 2-forms on $M$. This corresponds to the geometric moduli
space of deformations of a coassociative 4-cycle, determined by McLean in
\cite{McLean:1998}.

\subsection{Tangential modes}

For the tangential modes the $Q$- and $\Qd$-closure conditions are
just (\ref{eqn_Q_closure}) and (\ref{eqn_Qdag_closure}) with all
the indices replaced by worldvolume indices $a, b, c, \dots$.

\paragraph{Associative 3-cycles.}

On the 3-cycle it is convenient to represent the $Q$-closure
condition using the projector $\pi^2_{\bf 7}$ in terms of $\phi$
which gives

\begin{equation}
\phi_{ab}^{\phnt{ab}c}\phi_{c}^{\phnt{c}de}  \prt_{[d} A_{e]}   =
0 \; .
\end{equation}

\noindent When pulled back to the associative cycle, $\phi$ is
proportional to the volume form and so this is

\begin{equation}
\ep_{ab}^{\phnt{ab}c}\ep_{c}^{\phnt{c}de}  \prt_{[d} A_{e]}   = 0
\; ,
\end{equation}

\noindent which is just multiple copies of the equation $
\prt_{[d} A_{e]}   = 0$. Therefore any tangential deformation
corresponds to a flat connection on the 3-cycle.

Requiring the deformation $A_a \psi^a$ be also be $\Qd$-closed,
and hence a harmonic representative of $Q$-cohomology, implies
(\ref{eqn_Qdag_closure}), which can be viewed as enforcing a
covariant gauge condition.

Combined together this means that the $Q$-cohomology for tangential modes on $M$
is spanned by the space of gauge-inequivalent flat connections on $M$.  This
matches the result for Lagrangian submanifolds in the A-model and also
the results derived using $\kappa$-symmetry for physical branes in
\cite{Marino:1999af}.

\paragraph{Coassociative 4-cycles.}

On the 4-cycle it is easier to use the representation of
$Q$-closure

\begin{equation}
  \begin{split}
   \bigl( (\dl^a_c \dl^b_d - \dl^b_c \dl^a_d) + 24
  (*\phi)^{ab}_{\phnt{ab}cd} \bigr)
  \prt_{[a} A_{b]} \psi^c \psi^d  = 0 \; ,
 \end{split}
\end{equation}

\noindent in terms of the 4-form $*\phi$, which is now
proportional to the volume form on $M$. Defining $F_{ab} =
\prt_{[a} A_{b]}$ to be the field strength of the $U(1)$ gauge
field, the equation above implies

\begin{equation}
     ( *_4 F)_{ab} = 12 (*\phi)^{cd}_{\phnt{cd}ab}  F_{cd}  = - F_{ab} \; .
\end{equation}

\noindent Thus $F_{ab}$ is constrained to be anti-self-dual (ASD)
on $M$. Therefore any tangential deformation on the 4-cycle is
given by a gauge field with ASD field strength. Note an important
difference with the case of normal modes. In the latter case each
$\tht^I$ is mapped uniquely to a harmonic self-dual 2-form
$\Om_{ab}$ on $M$, so there are exactly $b^2_+ (M)$ such modes. In
this case however the tangential mode $A_a$ is the potential for a
gauge field with ASD field strength (i.e. an (anti-)instanton
configuration). Hence the tangential modes correspond to tangent
vectors on the moduli space of instanton configurations on $M$.

Again the condition $\nabla_a A^a = 0$ for $\Qd$-closure is simply
a gauge choice, implying that each $Q$-harmonic representative is
associated to a unique orbit of the gauge group (up to Gribov
ambiguity in the path integral). In fact, these harmonic
constraints $*_4 F = -F$, $d^\dagger A =0$ are precisely
(linearized versions of) the conditions cited in equation (5.22)
of \cite{Birmingham:1991ty} as defining the deformations of an
instanton configuration.

In physical string theory the anti-self-duality constraint on the field strength
of a coassociative brane has been determined in \cite{Marino:1999af} using
$\kappa$-symmetry of the DBI action.  In \cite{Leung:2002qs}, a topological
field theory is proposed on calibrated 4-cycles whose total moduli space is a
product of the moduli space of geometric deformations with the moduli space of
ASD connections on $M$. We will see shortly that this is indeed the worldvolume
theory on coassociative 4-cycles for the open $G_2$ string.

\section{Scattering amplitudes}

Before considering the nature of the worldvolume theory of the
calibrated 3- and 4-cycles, it will be useful to consider some
scattering amplitudes in the open $G_2$ theory, as these can be
compared with field theoretic scattering amplitudes and will help
constrain the interaction terms in the worldvolume action.  In
fact, as will be discussed in the next section, these interactions
can actually be related to string field theory, not just to
effective field theory, if one concedes that the $G_2$ string is
independent of its coupling constant, as argued in
\cite{deBoer:2005pt}.

\subsection{3-point amplitude}\label{subsec_3_pt}

The simplest amplitudes to calculate (and the only ones we will
need) are the 3-point functions of degree one fields $A_\mu
\psi^\mu$, which are essentially already calculated in
\cite{deBoer:2005pt}.  Introducing Chan-Paton factors into the
calculation performed there gives the 3-point function of three
ghost number one fields as

\begin{equation}
  \lm^3 \frac{3}{2} f_{jik} \int_Y \phi^{\al\bt\gm}(x)  A_\al^i(x)  A_\bt^j(x)
  A_\gm^k(x)\; ,
\end{equation}

\noindent where $f_{ijk}$ are the structure functions for the Lie
algebra of the gauge group $G$ and $\lm$ is the normalization of
the bulk-boundary 2-point function in the $G_2$ CFT (these are
generally not relevant and will not be treated with a great deal
of care).

\paragraph{Tangential modes.}

For an associative 3-cycle embedding $i : M \rightarrow Y$, we
have the relation $i^*(\phi) = \ep$, where $\ep$ is the volume
form on $M$. If we now consider the previous calculation but where
now the fields $\psi^\mu$ are restricted to be along the 3-brane
(so they have indices $a, b, c\dots$), we find that

\begin{equation}
  \braket{A A A} = \lm^3 \frac{3}{2} f_{jik} \int_M \ep^{abc}(x)  A_a^i(x)  A_b^j(x)
  A_c^k(x)\; .
\end{equation}

\noindent As will be discussed in the next section, this is an
interaction vertex for Chern-Simons theory, which is the part of
the effective worldvolume theory for the 3-cycle.

As mentioned in previous section, on a coassociative 4-cycle
$\phi^{abc} = 0$ so the 3-point function of tangential modes
vanishes.

\paragraph{Normal and mixed modes.}

We can now try to consider a mixture of normal or tangential modes
in the 3-point function.  The boundary conditions on the open
$G_2$ string, preserving the extended algebra on a 3-cycle, imply
\cite{Becker:1996ay} that only $\phi^{abc}$ and $\phi^{IJc}$ are
non-vanishing. Thus $\phi$ is only non-vanishing for an even
number of indices in Dirichlet directions, so we can only scatter
two normal modes and one tangential mode. This gives

\begin{equation}
  \braket{\tht \tht A} = \lm^3 \frac{3}{2} f_{jik} \int_M \phi^{IJc}(x)  \tht_I^i(x)  \tht_J^j(x)  A_c^k(x) \; .
\end{equation}

\noindent On a 4-cycle the non-vanishing components of $\phi$ have
an odd number of normal indices, and it is easy to see that the
only non-vanishing 3-point functions of degree one modes are

\begin{equation}
  \begin{split}
  \braket{\tht A A} = \lm^3 \frac{3}{2} f_{jik} \int_M \phi^{Iab}(x)
  \tht_I^i(x)  A_a^j(x)  A_b^k(x) \; , \\
  \braket{\tht \tht \tht} = \lm^3 \frac{3}{2} f_{jik} \int_M \phi^{IJK}(x)
  \tht_I^i(x)  \tht_J^j(x)  \tht_K^k(x) \; .  \\
  \end{split}
\end{equation}

\section{Worldvolume theories}\label{sec_worldvolume_theory}

We have already determined the BRST cohomology of normal and
tangential modes on 3- and 4-cycles.  These should be thought of
as marginal deformations of the theory preserving the twisting on
the worldsheet (by general arguments that map an element of BRST
cohomology to a descendant that can generate a deformation). When
considered from the spacetime perspective, the elements of BRST
cohomology should translate into spacetime fields and we expect
the BRST closure condition to correspond to the {\em linearized}
spacetime equations of motion.  This is true in physical string
theory and can be derived more rigorously via open string field
theory for topological strings, as will be reviewed below.

For the normal modes, the BRST cohomology condition can be
translated into constraints on deformations of the calibrated
submanifolds, such that these modes correspond to tangent vectors
on the moduli space of (co)associative cycles in the $G_2$
manifold.

For tangential modes, the BRST cohomology condition looks
different for the different cycles. On the 3-cycle, BRST closure
and co-closure of the tangential mode $A_a$ imply $dA = 0$ and
$d^\dagger A =0$, so that $A$ is a flat connection in a fixed
gauge, and we expect a gauge theory whose solutions correspond to
gauge-inequivalent flat connections. On the 4-cycle, BRST closure
and co-closure of $A_a$ imply

\begin{equation}\label{eqn_4cycle_tangent}
  *_4 dA = - dA \; , \qquad \qquad d^\dagger A = 0 \; .
\end{equation}

These equations are the linearization of the condition for a
variation of a gauge field to be a deformation of an instanton
solution (c.f. equation (5.50) in \cite{Birmingham:1991ty}). This
suggests, in analogy with the geometric moduli, that the theory on
the worldvolume should be a gauge theory extremizing on instantons
and that marginal tangential deformations of the worldsheet theory
should correspond to tangent vectors on the moduli space of
instantons.

In the case of both the 3- and 4-cycle, the worldvolume theory
will include contributions from the normal and tangential modes,
and so should result in a theory whose moduli space includes the
normal and tangential deformations that we have determined in
section \ref{subsec_moduli}. We also expect that the other
physical states, which are massless in the twisted theory, may
still play a role in the spacetime action even though they cannot
be used to generate boundary deformations of the CFT\footnote{Only
a ghost number one state has a 1-form descendant with ghost number
0; ghost number $p$ states have $p$-form descendants with ghost
number zero, so to preserve the ghost number in the worldsheet
action we would have to integrate them over a $p$-cycle on the
worldsheet.}, and hence are not moduli of the theory.

To determine the relevant spacetime actions and how the normal and tangential
moduli, as well as the higher ghost number fields, come into play we will start
by considering Witten's derivation of Chern-Simons theory from open string field
theory (OSFT).  We will find that by restricting our attention to tangential
modes on a calibrated 3-cycle we can re-derive Witten's Chern-Simons theory
simply by following the arguments of \cite{Witten:1992fb}.  We will then attempt
to generalize this derivation to include normal modes.  Their contribution is
expected to be related to the topological theories in \cite{Anguelova:2005cv,
Bonelli:2005rw}, whose actions also localize on the moduli space of associative
3-cycles (though, as we will see, this relation is mostly at the level of
equations of motion). Following a comment in \cite{Witten:1992fb}, we expect the
higher string modes to be related to additional fields generated by gauge-fixing
the CS action.  This is discussed in appendix \ref{app_ghosts}.

Once we have transplanted Witten's arguments for special
Lagrangian branes in a Calabi-Yau to associative branes in a $G_2$
manifold, we will apply them to branes wrapping coassociative
cycles and branes wrapping all of $Y$.

\subsection{Chern-Simons theory as a string theory}

In \cite{Witten:1992fb} Witten argues that the open A-model on
$T^*M$ reduces exactly to Chern-Simons theory on $M$, for any
3-manifold $M$. There are several arguments supporting this claim
and we will attempt to generalize them below to the $G_2$ case.
Before doing so, we first review them briefly.

The first argument concerns $Q$-invariance of a boundary term in
the string path integral.  In general the open string path
integral can be augmented by coupling to a \lq classical'
background gauge field.  This is done by including an additional
piece in the integrand of the path integral which is of the form

\begin{equation}\label{eqn_boundry_coupling}
  \tr P \exp\left( \oint_{\prt \Sg} X^*(A) \right) \; .
\end{equation}

\noindent Here $A$ is a (non-abelian) connection defined on the
brane $M$ and the term above is a Wilson loop for the pull-back of
this connection along the boundary of the worldsheet $\Sg$.
Requiring that this new term preserve the $Q$-invariance of the
action implies that the field strength $F = dA + A\wedge A$ must
vanish.  Hence open strings in the A-model can only couple to flat
connections.

To more rigorously establish that the relevant spacetime theory is
Chern-Simons theory, Witten considers the OSFT action

\begin{equation}\label{eqn_osft_action}
  \int \CA \star Q \CA + \frac{2}{3} \CA \star \CA \star \CA \; ,
\end{equation}

\noindent where $\CA$ is a functional of the open string modes
quantized on a fixed time slice, and $Q$ is the appropriate BRST
operator of the theory.  The integration measure is defined by the
path integral over the disc\footnote{There is a subtlety here.  In
OSFT for the bosonic string this measure involves gluing together
several discs using conformal transformations, but in the setting
of a topological theory all the states have conformal weight zero
under the twisted stress-tensor so they do not transform under
conformal transformations.}.  The linearized equations of motion
(coming from the quadratic part of the OSFT action) enforce the
requirement that physical states are BRST-closed on-shell:

\begin{equation}
Q \CA = 0 \; .
\end{equation}

\noindent In the large coupling constant limit ($t \ra \infty$) the
$Q$-cohomology can be studied by restricting to functionals $\CA$ that depend
only on the string zero-modes, $X^\mu_0$ and $\psi^\mu_0$. The BRST operator,
$Q$, acting on such states, reduces to the exterior derivative $d$ on $T^* M$
(which we can write in terms of the zero modes)

\begin{equation}
  d = dx^\mu \frac{\prt}{\prt x^\mu} = \psi^\mu_0 \frac{\prt}{\prt X^\mu_0} \; .
\end{equation}

\noindent Since the $t \ra \infty$ limit is exact in the A-model (modulo
world-sheet instantons which are not present when the target space is $T^*M$),
these identifications are not approximations but rather exact statements.  This
allows one to identify the string field action with Chern-Simons theory.

To make this identification one must identify the string field
$\CA$ with the target space gauge field $A_\mu(x) dx^\mu$.  The
general form for $\CA$ at large $t$ is given by the expansion

\begin{equation}\label{eqn_str_fld}
  \CA(X^\mu, \psi^\mu) = f(X_0) + A_\mu(X_0) \psi_0^\mu + \bt_{\mu\nu}(X_0)
  \psi_0^\mu \psi_0^\nu + C_{\mu\nu\rho}(X_0)\psi_0^\mu \psi_0^\nu
  \psi_0^\rho\; ,
\end{equation}

\noindent in 3 dimensions. The reason that $\CA$ reduces to
$A_\mu(X_0)\psi_0^\mu$ is simply that only ghost number one string
fields should be considered, and here ghost number coincides with
fermion number. Witten comments that it is possible to relate the
other terms in the expansion to ghost and anti-ghosts fields
derived from gauge-fixing CS theory \cite{Axelrod:1991vq}, or
alternatively gauge-fixing OSFT.  In appendix \ref{app_ghosts} we
will show that this is indeed the case when we repeat this
derivation on an associative cycle in a $G_2$ manifold.

Witten provides a final argument for CS theory as the string field
theory for the A-model, namely that the open string propagator on
the strip reduces to the CS propagator in the large $t$ limit.
This is essentially the statement that $\frac{b_0}{L_0} =
\frac{d^\dag}{\Box}$.  For the topological string, $b_0$ is
replaced by the superpartner of the stress-energy tensor in the
twisted theory (i.e. $Q^\dag$ in $T = \{Q, Q^\dag\}$).  In the
$G_2$ case this would be (tentatively) $\Gup$
\cite{deBoer:2005pt}.

We will now attempt to establish the validity of these arguments
for the open $G_2$ string ending on a calibrated 3-cycle.  Before
doing so we should mention that what was missing in this treatment
is a discussion of the normal modes on the brane.  It is not
immediately clear whether these modes modify the Chern-Simons
action on the special Lagrangian cycle (though one would imagine
they should in order to capture the dependence of the theory on
the geometric moduli of the brane).

\subsection{Chern-Simons theory on calibrated submanifolds}

If we consider only the tangential modes on a calibrated cycle
then the $Q$-closure conditions become (in the free field
approximation)

\begin{align}
  \prt_a f(X) = 0 \; ,\label{eqn_q_close_f}   \\
\ep^{abc}  \prt_{a} A_{b}   = 0 \; ,\nonumber \\
\ep^{abc} \prt_{a} \bt_{bc} = 0 \; ,\label{eqn_q_close_bt}
\end{align}

\noindent for the degree 0, 1, and 2 components of the string field.  Here we
have already used that $\phi_{abc} \propto \ep_{abc}$ on the 3-cycle.  This is
consistent with the notion that $Q = \Gdn = d$ in the large $t$ limit.  More
generally, the complex (\ref{eqn_g2_complex}), which encodes the BRST
cohomology, reduces, when restricted to the tangential directions on an
associative 3-cycle, to the de Rham complex so $Q = d$ and $Q^\dag = d^\dag$.

Recall, from the discussion in Section \ref{subsec_g2_hilb}, that,  in contrast
to the situation in the A-model, we do not have an explicit worldsheet action to
work with and hence do not have a Hamiltonian formulation which might directly
establish the $t$ invariance of the action.  Assuming this invariance
none-the-less, the equations above imply that the quadratic part of the string
field action reduces to the quadratic part of Chern-Simons theory.  That is, in
the large $t$ limit, the $Q$-closure constraint becomes the linearized CS
equation of motion.  Here we have also considered modes with fermion number
different from one; these will be discussed in appendix \ref{app_ghosts} in
relation to gauge-fixing Chern-Simons theory.

Also in this limit (of free string theory approximation), the
$\Qd$-closure constraints become

\begin{align}
\nabla_a A^a = 0 \; , \nonumber \\
\nabla_a \bt^{ab} = 0 \; . \label{eqn_qdag_close_bt}
\end{align}

\noindent The first term is just the gauge-choice $d^\dag A = 0$.  We will
discuss the spacetime interpretation of $\bt_{ab}$ in appendix \ref{app_ghosts}
and it will be clear why it satisfies this constraint. Let us now translate the
rest of Witten's arguments to the $G_2$ case.

The argument is essentially that open string field theory with the
action (\ref{eqn_osft_action}) reduces to Chern-Simons theory in
the large $t$ limit, if one restricts the string field to have
ghost number 1 (which, in the $G_2$ case, translates into fermion
number 1 because the ghost number is the grading for the
$Q$-cohomology, and that is given by the fermion number).  That
this holds for the kinetic term  follows because we have shown
that the linearized CS action is the same as the linearized
$Q$-closure condition.

For the interaction term this just follows from the fact that the 3-pt function
of the ghost number one parts of $\CA$ reduces to the wedge products of the Lie
algebra valued 1-forms, $A_a(x)dx^a$.  This is because, at large $t$, $\CA$
depends only on the zero modes so the ghost number one part has the form
$A_a(X_0) \psi^a_0$ which can be mapped to one-forms in spacetime.  We show in
section \ref{subsec_3_pt} that the 3-pt function of these modes is just the 3-pt
correlator of CS theory.

Witten also shows that the propagator of the OSFT reduces, in the
$t \ra \infty$ limit, to the CS propagator.  We will reproduce
this argument briefly here for the $G_2$ case.  A much more
complete treatment (of the analogous A/B-model argument) can be
found in section 4.2 of \cite{Witten:1992fb}. The open string
propagator is simply given by the partition function of a finite
strip, of length $T$ and width 1 with the standard metric

\begin{equation}
ds^2 = d\sg^2 + d\tau^2 \; .
\end{equation}

\noindent In OSFT the moduli space of open Riemann surfaces is
built by gluing such strips together.  The strip has one modulus,
namely its length, so in calculating the partition function, one
insertion of $\Gup$ folded against a Beltrami differential $\mu$
is required \cite{deBoer:2005pt}

\begin{equation}
\int d\sg d\tau \; \mu(\sg, \tau) G^\upar(\sg, \tau) \; .
\end{equation}

\noindent The Beltrami differential here is just given by a
change to the metric that changes the length of the strip and is given by a
function $f(\tau) = \dl T \cdot \dl(\tau - \tau_0)$ for any $\tau_0$ on the
strip.  Here $\dl T$ is the infinitesimal change in the length of the strip
generated by this differential. Thus the insertion becomes

\begin{equation}
\int d\sg d\tau \; \dl T \cdot \dl (\tau - \tau_0) G^\upar(\sg,
\tau) = \dl T \int d\sg  \;   G^\upar(\sg, \tau_0) \; .
\end{equation}

\noindent Because we have been working in the NS sector, the integral of the
current $G^\upar(z)$ around a contour (given by fixed $\tau_0$ which maps to a
half-circle in the complex plane) will just give a $\Gup$ insertion in the
world-sheet path integral, so its overall form is

\begin{equation}
  \int_0^\infty DT (\Gup) e^{-TL_0} = \frac{\Gup}{L_0} \; .
\end{equation}

\noindent By our previous identification of $\Gup$ with $d^\dag$ (this becomes
$d^\dag$ on $M$ for tangential modes) and $L_0$ with $\Box$ in the large $t$
limit, this becomes $\frac{d^\dag}{\Box}$ which is the CS propagator
\cite{Witten:1992fb}.  One should note that in the A-model this follows rather
directly but in the $G_2$ string it depends on the fact that $\phi_{abc} \propto
\ep_{abc}$ on the associative cycle (so, as previously mentioned, $Q=\check{D}$
reduces to $d$) and thus, in particular, might not hold on a coassociative
cycle.

There is a final argument one can make in favour of CS theory, though it is more
heuristic.  We want to argue, as Witten has, that coupling the worldsheet to a
classical background gauge field via a term such as (\ref{eqn_boundry_coupling})
requires this background to satisfy $F = 0$ which is the equation of motion for
Chern-Simons theory.

In \cite{deBoer:2005pt}, a heuristic version of the twisted $G_2$ action is
derived using the decomposition of worldsheet fermions into $\upar$ and $\dnar$
components, $\psi = \psi^\upar + \psi^\dnar$. This is heuristic because this
decomposition is essentially quantum and is not understood at the level of
classical fields.  Using this decomposition we can check Witten's argument for
the BRST-invariance of a boundary coupling to a classical configuration of the
gauge field

\begin{equation}\label{eqn_gauge_coupling}
   \tr P \exp\left( \oint_{\prt \Sg} A_\mu \prt_t X^\mu \right) \; .
\end{equation}

\noindent The variation of this factor in the partition function under  $[Q,
X^\mu] = \dl X^\mu$ is given by

\begin{equation}
  \tr P \oint_{\prt \Sg} \dl X^\mu \prt_t X^\nu F_{\mu\nu} d\tau \cdot
  \exp\left(
  \int_{\prt \Sg; \tau} A_\mu \prt_t X^\mu \right) \; ,
\end{equation}

\noindent where the contour in the exponent must start and end at the point
$\tau$ \cite{Witten:1992fb}.  To make this variation vanish requires that the
first term vanish and since \cite{deBoer:2005pt}

\begin{equation}
  \dl X^\mu = i \ep_L \psi^{\dnar\mu}_L + i \ep_R
  \psi^{\upar\mu}_R\; ,
\end{equation}

\noindent this implies that $F_{\mu\nu} = \prt_{[\mu} A_{\nu]}  + [A_\mu,
A_\nu] = 0$ for classical configurations of the background gauge field $A$.
This is, of course, the Chern-Simons equation of motion.

In the physical theory one could also couple to a term of the form
$C_{\mu\nu}\psi^\mu\psi^\nu$, but no such terms seem to effect the
derivation of $F = 0$ above in the A-model, because any such
coupling results in a variation which cannot cancel the
gauge-field coupling.

The only boundary term in a topological theory should be generated by the
descent procedure starting from a $Q$-closed ghost number one field whose
descendent is a ghost number zero one-form that is given by

\begin{equation}
  \{\Gup, A_\mu \psi^\mu\} = A_\mu \prt_t X^\mu + \pi^2_{\bf 14}(\prt_\mu A_\nu
  \psi^\mu\psi^\nu) \; .
\end{equation}

\noindent Both these terms have conformal weight 1 and, by virtue
of a standard descent argument, are $Q$-closed up to a total
derivative.  To apply Witten's argument here it is necessary to
understand why the second term cannot appear on the boundary. This
follows because we are considering modes tangential to an
associative cycle and one can check that on such a cycle
$\Lm^2 T^*M \sub \io^*(\Lm^2_7(Y))$ (here $\io: M \ra Y$ is the
embedding of the three cycle into the ambient $G_2$).

To derive the Chern-Simons action we have considered only the ghost number one
part of the string field $\CA$ as this is the standard prescription in OSFT.  In
some cases, however, it is desirable to consider the full expansion of $\CA$ and
include fields of all ghost number in the action.  This is because the higher
modes just play the role of ghosts in gauge-fixing the OSFT action
\cite{Thorn:1986qj}.  This is a special feature of Chern-Simons like theories
\cite{Axelrod:1991vq} and so will apply for all the brane theories that we
derive.  We include an appendix \ref{app_ghosts} describing the general form of
the gauge-fixed actions for these theories that we will need when we consider
their one-loop partition functions.

\subsection{Normal mode contributions}

In the previous section we argued that the tangential modes of the
$G_2$ worldsheet correspond to gauge fields in a CS theory on the
3-cycle and when higher string modes are included this becomes
gauge-fixed CS theory.

We are also interested in terms in the effective action that include the normal
modes.  The most direct way to to get at a normal mode action is to simply
expand the terms $\CA \star Q \CA $ and $\CA \star \CA \star \CA $ in the OSFT
action.  Ignoring the higher string modes, we have

\begin{equation}
  \begin{split}
    \CA &= A_a\psi^a + \tht_J \psi^J \; ,\\
    Q \CA &= \{Q, A_a \psi^a\} + \{Q, \tht_I \psi^I\} \\
    &= \phi_{IJc}\phi^{cde} \nabla_d A_e \psi^I \psi^J
      + \phi_{abc}\phi^{cde} \nabla_d A_e \psi^a \psi^b
      + \phi_{aIJ}\phi^{JbK} \nabla_b \tht_K \psi^a \psi^I \; .
  \end{split}
\end{equation}

\noindent Recall that the integration of expressions involving
string fields, $\CA$, in the OSFT action corresponds to evaluating
the correlator of the integrand, decomposed in individual string
modes on the disc.  In the $G_2$ string only certain combinations
of string modes will have a non-vanishing 3-pt function depending
on the conformal blocks the modes correspond to (see
\cite{deBoer:2005pt}). In our calculation of the 3-pt functions
above, this translates into non-vanishing 3-pt functions when we
can contract the spacetime indices of the string modes with the
3-form $\phi$.  From our previous calculation of three point
functions in sections \ref{subsec_3_pt} (see also 
App. \ref{subsec_ghost_correl}) we find the generic form of a 3-pt
function on the disc

\begin{equation}\label{eqn_gen_3pt}
  \begin{split}
    \braket{\lm \om} = \int_M \phi^{\mu\nu\rho} \tr \left( \lm_\mu \om_{\nu\rho} \right) \; ,\\
    \braket{\al \bt \gm} = \int_M \phi^{\mu\nu\rho} \tr \left( \al_\mu \bt_\nu
    \gm_\rho \right) \; ,
  \end{split}
\end{equation}

\noindent (where, e.g. $\om = 1/2 \, \om_{\mu\nu} (x) \psi^\mu \psi^\nu$). Doing this gives the following action

\begin{equation}\label{eqn_cs_normal_action}
  S_{\textrm{deg 1}} = \int_M
  \phi^{abc} \, \tr \left( A_a \nabla_b A_c + \frac{2}{3} A_a A_b A_c \right)
  + \phi^{IaJ} \, \tr \bigl( \tht_I ( \nabla_a \tht_J + [ A_a , \tht_J ] )
  \bigr)\; ,
\end{equation}

\noindent where the trace $\tr$ is over Lie algebra indices. The interaction terms can be calculated directly in string
perturbation theory by checking 3-pt disc amplitudes whereas the kinetic terms
coming from $\CA \star Q \CA$ vanish in perturbation theory because on-shell
string modes satisfy $Q \CA = 0$.  To determine them we either simply consider
all the terms of the correct degree in the string mode decomposition of $\CA
\star Q \CA$ or \lq formally' calculate 3-pt functions assuming the field $\CA$
is off-shell.  Both result in the same action and as a consistency check, the
linearized equations of motion for this action correspond to the BRST closure of the
string modes.
We have not been too careful with the coefficients in (\ref{eqn_cs_normal_action})
but this is because most coefficients either follow from gauge invariance or can
be absorbed into field redefinitions.

The equations of motion for this action are

\begin{align}
  \ep^{abc} F_{bc} &= \phi^{IaJ} [ \tht_I , \tht_J ] \; , \label{eqn_F_nonflat} \\
  \phi^{aIJ} \bigl( \nabla_a \tht_J + [ A_a , \tht_J ] \bigr) &= 0 \; . \label{eqn_gauged_normal}
\end{align}

\noindent In the abelian case this just reduces to $F=0$ and the
geometric constraint (\ref{eqn_norm_cond2}) on the normal modes
describing associative deformations.  In the non-abelian case this
is no longer true but of course in this setting we have lost the
simple association of $\tht_I$ with normal deformations of the
brane, as the string modes become matrix-valued.

At first glance the equations above look similar in form to the Seiberg-Witten
type equations (32) and (40) in \cite{akbulut-2004}. This reference is concerned
with resolving the singular structure of the moduli space of deformations of
associative submanifolds in a general $G_2$ manifold by considering a larger
space of deformations where one is allowed to also deform the induced connection
on the normal bundle to make the deformed submanifold associative. This amounts
to a choice of complex structure on the normal bundle, for each deformation of
the 3-submanifold, such that its reduced structure group $U(2) \subset SO(4)$ in
the $G_2$ manifold is compatible with the induced metric connection. This
additional topological restriction on the $G_2$ manifold is something we have
not assumed and indeed, for general gauge group, there is no obvious relation
between ({\ref{eqn_F_nonflat}}), ({\ref{eqn_gauged_normal}}) and the purely
geometric equations in \cite{akbulut-2004}\footnote{It is possible that the
$U(1)$ part of our gauge connection could be related to the $U(1) \subset U(2)$
part of the induced connection on the normal bundle with fixed complex structure
in \cite{akbulut-2004}.}.

\subsection{Anti-self-dual connections on coassociative submanifolds}

We expect that the worldvolume theory on the 4-cycle should have
equations of motion corresponding to the BRST closure of the
associated string modes.  Let us consider the following action

\begin{equation}\label{eqn_4cycle_actfull}
S[A,\theta ] = \int_M \phi^{Iab} \tr \bigl( \tht_I F_{ab} \bigr) +
\frac{2}{3} \, \phi^{IJK} \tr \bigl( \theta_I  \theta_J  \theta_K
\bigr) \; .
\end{equation}

As with the action on a 3-cycle we cannot directly check the quadratic terms by
considering a string correlator because the relevant correlators vanish for
on-shell states as dictated by the fact that the quadratic terms in the action
determine the BRST closure condition.  Rather, we can compare the linearized
equations of motion (generated purely by the quadratic terms) and the string
BRST closure condition and these should match.

The abelian $\tht_I$ equation of motion is now just $\phi^{Iab} F_{ab} =
0$, which implies anti-self-duality of $F$ and so matches the BRST
closure condition. The $A_a$ equation of motion is

\begin{equation}\label{eqn_gauged_normal2}
  \phi^{abI} D_b \theta_I = 0 \; ,
\end{equation}

\noindent where $D_a = \nabla_a + [ A_a , ]$ on $M$. This equation is more
conveniently expressed in terms of the self-dual 2-form $\om_{ab}
= \phi_{abI} \theta^I$ on $M$. At the linear level, the equation
above implies $\omega$ is co-closed, and hence also closed
since it is self-dual. Thus we have the correct linearized
condition for coassociative deformations found by McLean.

We can also consider the formal structure of the term $\CA \cdot Q
\CA$ in the OSFT action, letting $\CA$ go \lq off-shell', and
indeed we find matching.

As a further check we should compare the interaction term to
string scattering amplitudes.  The 3-pt function for a general
degree one vertex operator in the topological theory is given by

\begin{equation}
  \lm^3 \frac{3}{2} \int_Y \phi^{\al\bt\gm}(x) \tr \bigl( A_\al (x) A_\bt
  (x) A_\gm (x)  \bigr) \; .
\end{equation}

\noindent On the 4-cycle the only non-vanishing components of
$\phi$ must have an even number of tangential indices, which
implies the following non-vanishing amplitudes

\begin{equation}\label{eqn_3pt_4_cycle}
  \begin{split}
  \lm^3 \frac{3}{2} \int_M \phi^{Iab} \tr \bigl(  \tht_I A_a  A_b  \bigr) \; , \\
  \lm^3 \frac{3}{2} \int_M \phi^{IJK} \tr \bigl( \tht_I
   \tht_J  \tht_K  \bigr) \; .
  \end{split}
\end{equation}

The first line above corresponds to the cubic interaction $\tht A A$ in the
first term of (\ref{eqn_4cycle_actfull}) while second correlator in
(\ref{eqn_3pt_4_cycle}) implies the cubic vertex in the second term.  This last
term, of course, only corrects the non-abelian instanton equation of motion

\begin{equation}\label{eqn_FnonSD}
  \phi^{Iab} F_{ab} = - \phi^{IJK} [ \tht_J , \tht_K ] \; ,
\end{equation}

\noindent and so has no effect on the geometric interpretation in the abelian
case.

In \cite{Leung:2002qs} Leung proposes a 1-form on the space $\CC =
{\mbox{Map}}(M, Y) \times \CA(M)$ where $M$ is a 4-manifold, $Y$ is a $G_2$
7-manifold and $\CA(M)$ is the space of Hermitian connections on the gauge
bundle $E \ra M$ (with fibre $G$)

\begin{equation}
S(f, D_E)(v, B) = \int_M \tr \left( f^*(\io_v \phi) \wedge F_E +
f^*(\phi) \wedge B \right) \; .
\end{equation}

\noindent Here $(f, D_E) \in \CC$ and $(v, B) \in T_{(f, D_E)} \CC$ with $v$ a
section of $TY$, $B \in \Lambda^1(M, {\mbox{ad}} \, G)$ and $F_E$ the curvature
of $D_E$ (here $f$ is an element of ${\mbox{Map}}(M, Y)$ and should not be
confused with the $f$ used to denote the zero fermion component of the string
field). The one-form $S$ is invariant under diffeomorphisms of $M$ and its zeros
correspond to coassociative embeddings $f(M) \sub Y$ with anti-self-dual
connections on them.  This follows from the fact that $S$ must vanish when
evaluated on arbitrary vectors, $B$, implying $f^*(\phi) = 0$, and arbitrary $v$
implying that $F_E = -{*F}_E$.

To compare with our theory we do not want to consider the space of all such
maps, but only the local deformations of a given coassociative
$f(M)$ in $Y$, so we only consider fluctuations around a fixed
coassociative submanifold. Thus we will take $f$ to be a
coassociative embedding implying that the second term in the
action above vanishes and ${*\phi}$ defines the volume form on the
embedded coassociative 4-cycle $f(M)$. Thus, we rewrite Leung's
functional to generate the following action
functional\footnote{More precisely, Leung's one-form, $\Phi_0$,
descends to a closed one-form on the space \lq
$\CC/\textrm{Diffeo(M)}$' and this form is locally the derivative
of a functional, $\mathcal{F}$, whose critical points are zeros of
$\Phi_0$.  Our action is most closely related to this functional.}

\begin{equation}\label{eqn_4cycle_action}
S^0[A , \theta ] = \int_M \tr \left( f^*(\io_\tht \phi) \wedge F
\right)
 = \int_M \phi^{Iab} \tr \bigl( \tht_I \left( \prt_a A_b +  A_a  A_b  \right)
 \bigr)\; ,
\end{equation}

\noindent using the identity $\epsilon_{abcd} \, \phi^{cdI} = 2 \,
\phi_{ab}^{\;\;\;\; I}$ on the coassociative cycle.  

Thus we see that the open $G_2$ string has reproduced the action Leung suggested
in order to study SYZ in the $G_2$ setting and it has also introduced an
additional term that is not present in Leung's action.

\subsection{Seven-cycle worldvolume theory}

As in physical string theory, it is natural to expect the 3- and 4-cycle theory
to look like the dimensional reduction of a theory on the whole 7-manifold
(which is trivially calibrated by its volume form $\phi \wedge *\phi$). Lee et
al \cite{Lee:2002fa}, who propose theories closely related to our 3- and 4-cycle
theories, claim that this theory should be related to (deformed)
Donaldson-Thomas theory \cite{Donaldson:1996kp}.

The 7-cycle theory can be determined exactly the same way as the 3- and 4- cycle
theory. For the interaction term we just calculate the 3-pt functions of the
(ghost number one) terms in  $\braket{\CA \star \CA \star \CA}$ given by
(\ref{eqn_gen_3pt}).  The kinetic terms, defining the linearized equations of
motion, should correspond to $Q \CA = 0$ and they should match $\CA \star Q
\CA$.

This gives the following action

\begin{equation}\label{sevenwv}
  \begin{split}
  S &= \int_Y \phi^{\mu\nu\rh} \tr \left( A_\mu \prt_\nu A_\rh + \frac{2}{3} A_\mu A_\nu A_\rh \right)
    = \int_Y *\phi \wedge CS_3(A) \; .
  \end{split}
\end{equation}

\noindent The equation of motion for this action is

\begin{equation}
  *\phi \wedge F = 0 \; .
\end{equation}

\noindent This is one of the equations in \cite{Donaldson:1996kp} where it is
argued to be associated with the 7-dimensional generalization of Chern-Simons
theory.  In the abelian theory this equation of motion is simply $\check{D}A =
0$ which has no global solutions which are not exact (i.e. $A = df$) because
$H^1_7(Y) = 0$ for $G_2$ manifolds.  Of course, as a gauge field $A$ need not be
a global one form and then this result no longer applies.  This is similar to
the situation one finds for Chern-Simons theory on a simply connected manifold.

Note that for the action (\ref{sevenwv}) to be gauge invariant under
large gauge transformations $*\phi$ must actually be an integral cohomology
class.  A similar issue arises in holomorphic Chern-Simons theory as mentioned
by Nekrasov in \cite{Nekrasov:2005bb} but, as the three-form $\Om$ is
holomorphic, it is not clear that it can always be normalized to be integral.
Nekrasov notes, however, that the integrality condition is precisely the
condition on the complex moduli of the CY to be solutions of the attractor
equations.  It would be interesting to understand if the integrality of $*\phi$
has a similar interpretation.

 In \cite{Lee:2002fa} the authors want to consider solutions to the deformed
 Donaldson-Thomas equation

\begin{equation}
  \label{}
  *\phi \wedge F = \frac{1}{6} F \wedge F \wedge F \; ,
\end{equation}

\noindent which would involve adding a term $CS_7(A)$ to the Lagrangian above.
It is not at all clear why such a term would appear in OSFT but in Section
\ref{sec_quant} we see that such a term does emerge in a rather interesting way
when quantizing this theory.

\subsection{Dimensional reduction, A- and B-branes}

Reducing the open topological $G_2$ string on $CY_3 \times S^1$ gives rise to both special Lagrangian A-branes and holomorphic B-branes on $CY_3$.  This
follows from the decomposition of $\phi$ and $*\phi$ in terms of the holomorphic
3-form and K\"{a}hler form on $CY_3$ (see appendix \ref{app_conventions}).  The A-branes
arise when reducing the associative 3-cycle action (\ref{eqn_deg2_3cycle}) in the
normal direction. The resulting action

\begin{equation}
\int_{M} \epsilon^{abc} \tr \left( A_a \nabla_b A_c + \frac{2}{3}
A_a A_b A_c \right) +\rho^{aij} \tr \bigl( \theta_i ( \nabla_a
\theta_j + [ A_a , \theta_j ] ) \bigr) \; ,
\end{equation}

\noindent is the real part of complex Chern-Simons
theory, where the indices $a,b,c=1,2,3$ are in the SLag while $i,j=4,5,6$ are in the
normal direction. The normal modes appear quadratically and can be integrated
out (see Section \ref{sec_quant} for a discussion of this issue on an associated
cycle).

Similarly we can reduce the 4-cycle action (\ref{eqn_4cycle_actfull}) in the
tangential direction. This is again a special Lagrangian brane in $CY_3$ but now
calibrated by $\hrho$ instead of $\rho$, and the worldvolume action is given by the
imaginary part of complex Chern-Simons theory
\begin{equation}
\int_{M} \rho^{iab} \tr(\theta_i F_{ab}) + {2 \over 3} \rho^{ijk}
\tr(\theta_i \theta_j \theta_k) \; ,
\end{equation}
with the additional constraint $D_a \theta_i =0$ for the
normal modes.

The B-branes are simplest to find starting from the 7-cycle worldvolume
theory (\ref{sevenwv}) and reducing on the $CY_3$. We find

\begin{equation}
  \begin{split}
S &= \int_{CY_3} { \hat{\rho} \wedge CS(A) +  k \wedge k \wedge \tr ( \lm F ) } \\
&= {1  \over 2i} \int_{CY_3} \Omega \wedge CS(A) - {1 \over 2i}
\int_{CY_3} {\bar \Omega} \wedge {CS(\bar{A})} + \int_{CY_3} k
\wedge k \wedge \tr ( \lm F ) \; ,
  \end{split}
\end{equation}

\noindent where $*\phi = {\hat \rho} \wedge dt  + {1 \over 2} k \wedge k$, $t$
parametrizes the circle direction, $\Omega = \rho + i {\hat \rho}$ is the
holomorphic 3-form of the Calabi-Yau, and $\lm= A_t$ is the scalar component of
the gauge field in the reduction. The action is the sum of B-model 6-brane and
${\bar {\mbox{B}}}$-model 6-brane actions (the appearance of the imaginary part of the
holomorphic 3-form rather than the real part is just a matter of convention).
The extra term in the action comes with the Lagrange multiplier $\lm$, and so
it expresses the constraint
$$
k \wedge k \wedge F =0 \; .
$$
This extra condition is related to stability of the brane (complexifies the
$U(N)$ symmetry).  Lower-dimensional 4-branes and 2-branes then follow by
further dimensional reduction, where again we obtain B- and ${\bar {\mbox{B}}}$-model
actions together with a stability condition.

It is remarkable that like the closed topological M-theory, the open topological
string also contains the A and B+${\bar {\mbox{B}}}$ models. Perturbatively the B+${\bar {\mbox{B}}}$-models are decoupled, and it would be interesting to understand if there is
a non-perturbative coupling between them.

%%%%%%%%%%%%%%%%%%%%%%%%%%%%%%%%%%%%%%%%%%%%%%%%%%%%%%%%%%%%%%%%%%%%%%%%%%%%%%%%%%%

\section{Gauge-fixing and quantization}\label{sec_quant}

Let us now consider the full expansion of the OSFT action without any constraint
on the ghost number of the fields. As found in appendix \ref{app_ghost_7cycle},
this gives the following expression for the action in seven dimensions

\begin{equation}\label{sevenwv_gh}
  \begin{split}
  S_{(7)} &= \int_Y \phi^{\mu\nu\rh} \, \tr \left(A_\mu \prt_\nu A_\rh + \frac{2}{3} A_\mu A_\nu A_\rh +
  \bt_{\mu\nu} \prt_\rh f + \bt_{\mu\nu} [ A_\rh ,f ] + \frac{1}{2} C_{\mu\nu\rh} \{ f , f \} \right)\\
    &= \int_Y *\phi \wedge \tr \left( A \wedge dA + \frac{2}{3} A \wedge A
	\wedge A  + \bt \wedge D f + \frac{1}{2} C \{ f , f\}
    \right) \; ,
  \end{split}
\end{equation}

\noindent where $f \in \Lambda^0_{\bf 1}$, $\beta \in \Lambda^2_{\bf 7}$ and $C
\in \Lambda^3_{\bf 1}$ are respectively the degree zero, two and three modes of
the string field $\CA$ in the adjoint representation of the gauge group and $D =
d + A$ is the gauge-covariant derivative. The purely bosonic (i.e. ghost number
one field) part of the action above has appeared (in conjunction with additional
bosonic terms) in topological quantum field theories studied in
\cite{Baulieu:1997jx} and \cite{Acharya:1997fos}. The interpretation of the
action above in terms of the Batalin-Vilkovisky antifield formalism is detailed
in appendix \ref{app_ghost_7cycle}.

\subsection{Weak coupling limit}

To help us understand the structure of the gauge theories we have
found for open strings ending on (co)associative calibrated
branes, we are more interested in the quantization of the
quadratic part of the non-linear action $S [A] = \int_Y *\phi
\wedge CS(A)$, expanded around solutions of the classical
equations of motion

\begin{equation}
  *\phi \wedge F = 0 \; .
\end{equation}

The partition function of this simplified theory corresponds to a
stationary phase approximation of the full theory in the weak
coupling limit. For the gauge theory on associative 3-cycles, we
will investigate how the normal modes modify the corresponding
calculation done by Witten \cite{Witten:1988hf} for pure
Chern-Simons theory.

The equation $*\phi \wedge F =0$ has been considered already in
\cite{Donaldson:1996kp} where it is argued to be the 7-dimensional
generalization of Chern-Simons theory that might provide an analog
of Casson/Floer theory for 7-manifolds. It is related to an
instanton equation for a gauge field on the $Spin(7)$ 8-manifold
$Y \times \R$. This relationship is directly analogous to the way
solutions of the Chern-Simons equation of motion $F=0$ on a
3-manifold $M$ correspond to critical points of the gradient flow
equations coming from the instanton equations $F=*F$ on $M \times
\R$. This fact will be important when we come to consider the
non-trivial phase factor in the path integral of this gauge
theory.

Expanding $S[A]$, for $A = A^0 + B$, to quadratic order in $B$
around a classical solution $A^0$ gives

\begin{equation}
  S [A] = S [ A^0 ] + \int_Y *\phi \wedge \tr \left( B
  \wedge DB
  \right) \; ,
\end{equation}

\noindent where $D = d + [A^0,]$ is here with respect to the background gauge
field solving $\phi^{\mu\nu\rho} F^0_{\nu\rho} = 0$. The linear term is of
course absent since it gives the $A^0$ equation of motion. Performing the BV
analysis of the quadratic action $S_{cl} [B] = \int_Y *\phi \wedge \tr ( B
\wedge DB )$ is straightforward and is given in appendix \ref{app_ghost_7cycle}
(it is also related to a linearization of the structure described for the full
theory in appendix \ref{app_ghost_7cycle}).

The resulting gauge-fixed action takes the familiar form

\begin{equation}
\int_Y *\phi \wedge \tr \left( B \wedge DB \right) + \tr \left(
\varphi D^\mu B_\mu + {\bar c} D^\mu D_\mu c \right) \; ,
\end{equation}

\noindent with $\varphi$ acting as Lagrange multiplier imposing the
gauge-fixing constraint in the action while ${\bar c}$, $c$
correspond to the fermions from the Faddeev-Popov determinant.

Formally the analysis of this gauge theory in 7 dimensions has
been almost identical to Witten's analysis of pure Chern-Simons in
3 dimensions. Indeed we can also use Schwarz's method of
evaluating the partition function for degenerate quadratic
classical actions to obtain the contribution

\begin{equation}
{\mbox{exp}}(ik S[A^0]) \; \frac{{\mbox{det}}(D_\mu
D^\mu)}{\sqrt{{\mbox{det}}(L)}} \; ,
\end{equation}

\noindent to the partition function of $ik \int_Y *\phi \wedge CS(A)$ (in the
weak coupling limit of large $k$) coming from a given gauge-equivalence class of
solutions $A^0$ of $*\phi \wedge F =0$.  We should stress that the structure of
the moduli space of solutions to $*\phi \wedge F =0$ is not understood so well
as that for flat connections in 3 dimensions. Witten \cite{Witten:1988hf}
restricts attention to Chern-Simons theory on 3-manifolds $M$ with the property
that the moduli space of flat connections, determined by equivalence classes of
homomorphisms from $\pi_1 (M)$ to the gauge group $G$, be finite. We do not know
whether one can take the moduli space of gauge-inequivalent solutions of $*\phi
\wedge F =0$ to be zero-dimensional by suitable choice of $G_2$ manifold $Y$.
Thus we cannot say whether the partition function can be expressed as a finite
sum over contributions of the form above.

The operator appearing in the denominator above is defined $L = *(*\phi \wedge
D) + D*$ and is understood as an antisymmetric 8x8 matrix of linear differential
operators mapping $\Lambda^1_{\bf 7} \oplus \Lambda^7_{\bf 1}$ to itself. It
follows by collecting $B_\mu$ and $\varphi$ in the first two terms in the
gauge-fixed quadratic action into an 8-vector. One can check that this
definition implies $L$ is elliptic and self-adjoint. It seems the natural
generalisation of the elliptic self-adjoint operator $L_- = *D + D*$ (restricted
to forms of odd degree in 3 dimensions) used by Witten in \cite{Witten:1988hf}
\footnote{In both dimensions 3 and 7, the addition of the $D*$ term in $L$ is
essential in order for it to be elliptic. This is simply because without it the
Pfaffian of the corresponding antisymmetric symbol matrices of odd rank would
vanish identically and so there could exist no inverse. The way of understanding
the need for ellipticity in physics terms is that we require the kinetic
operator in the quadratic action to be the inverse propagator. The propagator
only exists for the gauge-fixed action.}
. Another technical point we are
overlooking is whether $L = *(*\phi \wedge D) + D*$ is a regular operator. We
need not get into the precise definition, sufficed to say that regularity of an
operator guarantees one has a precise definition of its determinant in terms of
regularised zeta functions.

As explained in \cite{Witten:1988hf}, the contribution to the
partition function of Chern-Simons theory in 3 dimensions around a
given flat connection at weak coupling is closely related to the
partition function of an abelian 1-form gauge theory in 3
dimensions, which has been explicitly calculated by Schwarz and
shown to give the Ray-Singer analytic torsion of the de Rham
complex of the 3-manifold, and is thus a topological invariant.
However, this relation to Ray-Singer torsion is generally only
guaranteed for topological actions of the form $\int \omega \wedge
d \omega$, where $\omega$ is a bosonic/fermionic $p$-form of
odd/even degree in $(2p+1)$ dimensions. Thus we should not expect
the partition function of the 7-dimensional quadratic theory above
to be obviously related to Ray-Singer torsion. On the other hand,
since we are still in odd dimension, a theorem of Schwarz
\cite{Schwarz:2001sf} does suggest the partition function for this
gauge theory should be a topological invariant. In fact this
statement is only true modulo possible obstructions related to
non-trivial phase factors that we will now discuss.

\subsection{Phase of the determinant}

An important subtlety in both 3 and 7 dimensions is the role of
the phase of the determinant of the operator $L$. The theories
described by Schwarz are insensitive to this since they compute
absolute values of ratios of determinants of elliptic operators.
The Laplacian $D_\mu D^\mu$ appearing in the numerator is real and
positive-definite so there is no possible phase coming from its
determinant. We will now investigate the structure of this phase
for the 7-dimensional theory.

The expression for the phase in terms of the Atiyah-Patodi-Singer
$\eta$-invariant follows in the same way in both 3 and 7
dimensions; as the limit of a series in powers of the non-zero
eigenvalues $\lambda_i$ of the operator $L$ (at a given background
solution $A^0$ of $*\phi \wedge F =0$). In particular, as in
\cite{Witten:1988hf}, we find

\begin{equation}
  \frac{1}{\sqrt{\det(L)}} = \frac{1}{|\sqrt{\det(L)}|} \, \exp \left( \frac{i\pi}{2} \eta_L(A^0)
  \right)\; ,
\end{equation}

\noindent where

\begin{equation}
  \eta_L (A^0) = \frac{1}{2} \, \underset{s \rightarrow 0}{\mbox{lim}} \sum_i {\mbox{sign}}
  \lambda_i \, | \lambda_i |^{-s} \; ,
\end{equation}

\noindent denotes the $\eta$-invariant for the elliptic operator $L$ at
solution $A^0$.

In 3 dimensions Witten \cite{Witten:1988hf} uses the
Atiyah-Patodi-Singer index theorem for the classical twisted spin
complex ($L_-$ can be interpreted as a twisted Dirac operator) to
compute the difference of $\eta$-invariants between two flat
connections, $A = A^0$ and $A=0$, to be proportional to the
Chern-Simons action $\int_M CS(A^0)$ itself at $A^0$. The
proportionality factor is the dual Coxeter number $h (G)$ of the
gauge group $G$. This has the beautiful interpretation of the
level shift $k \rightarrow k + h (G)$ in the quantum
Chern-Simons action, that one also observes for current algebras
of conformal field theories in 2 dimensions.

The identification of $L_- = *D+D*$ in 3 dimensions with a twisted
Dirac operator follows by collecting the differential operators in
$L_-$ into a 4x4 antisymmetric matrix acting on the 4-dimensional
vector space $\Lambda^1 \oplus \Lambda^3$. This allows one to
write $L_- = \gamma^a  D_a$ in terms of the 3 4x4 antisymmetric
matrices $\gamma_a$, with components $( \gamma_a )_{bc} = -
\epsilon_{abc}$, $( \gamma_a )_{b4} = - \delta_{ab}$. These
matrices generate a subgroup $SU(2) \subset SO(4)$ and in an
appropriate basis can be written $\Gamma_a = i \sigma_2 \otimes
\sigma_a$ (in terms of Pauli matrices $\sigma_a$). Together with
$\Gamma_4 = i \sigma_1 \otimes 1$, they generate a representation
of the Clifford algebra acting on Dirac spinors in 4 dimensions.
By constructing the interpolating gauge field $A(t)$, for $t \in
[0,1]$ on $M\times [0,1]$ between 2 flat gauge fields $A(1)=A^1$
and $A(0)=A^0$ on $M$, this provides a suitable lift of $L_-$ on
$M$ to the twisted Dirac operator ${\tilde L}_- = \Gamma^a D_a
(A(t)) + \Gamma^4 \partial_t$ on $M\times [0,1]$. It is the
Atiyah-Patodi-Singer index theorem for ${\tilde L}_-$ that allows
Witten to compute the change in $\eta_{L_-}$ between 2 flat
connections.

We will now show that a similar structure follows for $L = *(*\phi
\wedge D) + D*$ in 7 dimensions. Again collecting the differential
operators in $L$ into an 8x8 antisymmetric matrix acting on the
8-dimensional vector space $\Lambda^1 \oplus \Lambda^7$ allows one
to express $L = \gamma^\mu  D_\mu$ in terms of the 7 8x8
antisymmetric matrices $\gamma_\mu$, with components $( \gamma_\mu
)_{\nu\rho} = - \phi_{\mu\nu\rho}$, $( \gamma_\mu )_{\nu 8} = -
\delta_{\mu\nu}$. It should be noted that the sub-matrices $(
\gamma_\mu )_{\nu\rho}$ do not form the adjoint representation of
the imaginary octonions despite the fact that they are identical
to the structure constants of this algebra. This is simply because
the octonions are not associative. This is to be contrasted with
the submatrices $( \gamma_a )_{bc}$ in 3 dimensions which give the
adjoint representation of the imaginary quaternions (i.e. the Lie
algebra of $SU(2)$). Nonetheless, together with $\gamma_8 = i 1$,
the full 8x8 matrices $\gamma_\mu$ generate a representation of
the Clifford algebra acting on Weyl spinors in 8 dimensions. The
corresponding action on Dirac spinors in 8 dimensions can be
expressed in terms of the 16x16 anti-Hermitian matrices
$\Gamma_\mu = \sigma_2 \otimes \gamma_\mu$, $\Gamma_8 = i \sigma_1
\otimes 1$. Thus by constructing the interpolating gauge field
$A(t)$ on $Y\times [0,1]$ between 2 solutions $A(1)=A^1$ and
$A(0)=A^0$ of $*\phi \wedge F =0$ on $Y$ we have a suitable lift
of $L$ on the $G_2$ manifold $Y$ to the twisted Dirac operator
${\tilde L} = \Gamma^\mu D_\mu (A(t)) + \Gamma^8 \partial_t$ on $Y
\times [0,1]$.

Before obtaining the change in $\eta_L$ from the
Atiyah-Patodi-Singer index theorem for ${\tilde L}$, it may be
illuminating to make a brief digression explaining how this lift
of $L$ is related to the elliptic complex

\begin{equation}
  0 \longrightarrow {\mbox{ad}}\, G \otimes \Lambda^0 \overset{D}{\longrightarrow} {\mbox{ad}}\, G \otimes \Lambda^1
  \overset{\frac{1}{4}(1-*\Psi \wedge) D}{\longrightarrow}
  {\mbox{ad}}\, G \otimes \Lambda^2_{\bf 7} \longrightarrow 0 \; ,
\end{equation}

\noindent on an 8-manifold $X$ of $Spin(7)$ holonomy, with Cayley
4-form $\Psi = *\Psi$, when $X = Y \times [0,1]$. This complex has
been used in the study of 8-dimensional topological quantum field
theories in \cite{Baulieu:1997jx}. The operator $\pi^2_{\bf 7} =
\frac{1}{4}(1-*\Psi \wedge)$ projects the 2-form in 8 dimensions
onto the 7-dimensional irreducible representation of $Spin(7)$.
The adjoint operators mapping to the left of the complex are
$D^\dagger$. As noted by Donaldson and Thomas, solutions of $*\phi
\wedge F =0$ on the $G_2$ manifold $Y$ correspond to fixed points
of the gradient flow from the $Spin(7)$ instanton equation $*F =
\Psi\wedge F$ on $Y \times \R$ (i.e. elements of the kernel of
$\pi^2_{\bf 7} D$).

The relation of this complex to the twisted spin complex for ${\tilde L}$
follows by observing the isomorphisms ${\sf S}_+ = \Lambda^0_{\bf 1} \oplus
\Lambda^2_{\bf 7}$ and ${\sf S}_- = \Lambda^1_{\bf 8}$ for the positive and
negative chirality spin bundles ${\sf S}_{\pm}$ on a $Spin(7)$ manifold (using
the conventions of \cite{Acharya:1997fos} where the $Spin(7)$-invariant spinor
$\theta \in {\sf S}_+$). The explicit isomorphisms following from Fierz
identities give $\psi_+ = \eta \theta - \frac{1}{4} \chi_{MN} \Gamma^{MN}
\theta$ and $\psi_- = - \psi_M \Gamma^M \theta$ ($M,N=1,...,8$) for any
$\psi_\pm \in {\sf S}_{\pm}$, where $\eta = \theta^t \psi_+$ is a scalar,
$\chi_{MN} = \frac{1}{2} \theta^t \Gamma_{MN} \psi_+$ is a 2-form obeying the
identity $\pi^2_{\bf 7} \chi = \chi$ and $\psi_M = \theta^t \Gamma_M \psi_-$ is
a 1-form. The action of the twisted Dirac operator $\Gamma^M D_M : {\sf S}_-
\rightarrow {\sf S}_+$ on these expressions gives $\Gamma^M D_M \psi_- = ( D^M
\psi_M ) \theta - ( \pi^2_{\bf 7} D \psi )^{MN} \Gamma_{MN} \theta$ hence
equating $\Gamma^M D_M$ acting on ${\sf S}_-$ with $\pi^2_{\bf 7} D + D^\dagger$
acting on $\Lambda^1_{\bf 8}$ in the complex above.  This is consistent with the
reduction of the lifted ${\tilde L}$ on $Y \times [0,1]$ to $L=*(*\phi \wedge D)
+ D*$ on $Y$.  Using this identification, one can check that the index of the
whole $Spin(7)$ complex above is identical to that for the twisted Dirac
operator on a $Spin(7)$ manifold.

This identification has been used by Reyes-Carri\'{o}n \cite{ReyesCarrion:1998si} to calculate the Atiyah-Singer index

\begin{equation}
  \int_{X} ch({\mbox{ad}}\, G) \hat{A}(TX) = \int_X {\mbox{dim}} (G) \, {\hat A}_2 (TX) +
  \frac{1}{24} \left( p_1 (TX) \wedge c_2 ({\mbox{ad}}\, G) + 2 \, (c_2({\mbox{ad}}\, G))^2 -4\, c_4({\mbox{ad}}\, G)
  \right)\; ,
\end{equation}

\noindent of the $Spin(7)$ complex above, on a closed $Spin(7)$ 8-manifold $X$.
The A-roof genus $\int_X {\hat A}_2$ here corresponds to the number of parallel
spinors on $X$ and so equals 1 if the holonomy is exactly $Spin(7)$ (and not a
subgroup thereof). For convenience, it is assumed in the formula above that the
gauge group is chosen such that the Chern classes $c_1 ({\mbox{ad}}\, G)$ and
$c_3 ({\mbox{ad}}\, G)$ both vanish (e.g. for $G = SU(N)$).

Consider now $X = Y \times [0,1]$ where $A(t)$ interpolates
between two solutions $A= A^0$ and $A=0$ of $*\phi \wedge F =0$ on
$Y$. The Atiyah-Patodi-Singer index theorem for ${\tilde L}$ is

\begin{equation}
  {\mbox{ind}} ( {\tilde L} ) = \int_{Y \times [0,1]} ch({\mbox{ad}}\, G) \hat{A}(T(Y \times [0,1])) -
  \frac{1}{2} [\eta_L (A^{0}) - \eta_L (0)] \; .
\end{equation}

The bulk integral can be evaluated using the Reyes-Carri\'{o}n
result on $X = Y\times [0,1]$. This is equal to the continuous
part of $\frac{1}{2} [\eta_L (A^{0}) - \eta_L (0)]$ and is given
by

\begin{equation}
  {\mbox{dim}} (G) + \frac{1}{24} \left( \frac{1}{2\pi} \right)^4 \int_Y
  \left[ - \frac{1}{2} \tr (R \wedge R) \wedge CS_3( A^0 ) + CS_7 ( A^0 )
  \right]\; ,
\end{equation}

\noindent as an integral over the $G_2$ manifold $Y$ with Riemann curvature $R$.
The Chern-Simons forms are

\begin{align}
  CS_3 (A) &= \tr \left( A \wedge dA + \frac{2}{3} \, A^3 \right) \; , \nonumber \\
  CS_7 (A) &= \tr \left( A \wedge (dA)^3 + \frac{16}{5} \, A^3 \wedge (dA)^2 +
  \frac{4}{5}  A^2 \wedge dA \wedge A \wedge dA +
    2\, A^5 \wedge dA + \frac{4}{7} \, A^7 \right) \; .
\end{align}

In general $\frac{1}{2} [\eta_L (A^{0}) - \eta_L (0)]$ can also
have a discontinuous contribution, corresponding to the spectral
flow of $L$, and is equal to (minus) the index of the lifted
operator ${\tilde L}$ itself. This has the effect of shifting the
continuous part of $\frac{1}{2} [\eta_L (A^{0}) - \eta_L (0)]$ by
$\pm 1$ if the eigenvalues $\lambda_i (t)$ of $L(A(t))$
(understood as a function of $t$) change sign when $t$ is varied
between 0 and 1 (a +1 shift corresponds to a change $\lambda_i <0$
to $\lambda_i >0$).

The addition of \lq constant' terms (that do not depend on the
particular choice of solutions $A^1$ and $A^0$) to $\frac{1}{2}
[\eta (A^{1}) - \eta (A^{0})]$ will have a trivial effect that can
be factored out of the overall phase structure of the theory and
ignored. Thus the effect of the spectral flow of a given operator
can only be ignored if it is a constant in this sense. This is the
case for Witten's analysis of $L_-$ in 3 dimensions. This is
obviously also true for the constant ${\mbox{dim}}(G)$ in the
change in the $\eta$-invariant above. It is not clear to us
whether the effect of the spectral flow of $L$ in 7 dimensions
will be significant and we will overlook this subtlety here.

Therefore it is clear that the phase structure of the
7-dimensional theory is much more complicated than just the level
shift that occurs in 3 dimensions. Nonetheless, let us examine
some of the terms in $\frac{1}{2} [\eta_L (A^{0}) - \eta_L (0)]$ in a bit more detail.

The term $\tr (R \wedge R)$, proportional to the first Pontrjagin
class of $Y$, which ordinarily can be a general element of
$H^4(Y,\R )$, is here somewhat constrained due to the fact that
$Y$ must have holonomy in $G_2$. In particular, this constrains
the curvature such that $\pi^2_{\bf 7} R =0$ (or
$R_{\mu\nu\alpha\beta} \phi^{\alpha\beta\gamma} =0$ in components)
so that the holonomy algebra is contained in the Lie algebra of
$G_2$. Decomposing

\begin{equation}
  H^4(Y,\R ) = H^4_{\bf 1} (Y,\R ) \oplus H^4_{\bf 7} (Y,\R )
  \oplus H^4_{\bf 27} (Y,\R ) \; ,
\end{equation}

\noindent into irreducible representations of $G_2$, one can check that the
constraint above implies $\tr (R \wedge R)$ has no ${\bf 7}$ part.  (This also
follows from lemma 1.1.2 in \cite{Joyce:1996}, although only compact
manifolds with full $G_2$ holonomy are considered there and so one has the
stronger constraint $b^4_{\bf 7} =0$ which we need not assume here.)

The cohomology group $H^4_{\bf 1} (Y,\R ) = \R$ has a very simple
structure, being spanned by constant multiples of the harmonic
4-form $*\phi$. Moreover, one can prove the identity $\tr (R
\wedge R) \wedge \phi = - |R|^2 \, {\mbox{vol}}$ implying the
constant multiplying the ${\bf 1}$ part of the first Pontrjagin
class is negative definite and vanishes only if the $G_2$ metric
is flat (this also follows from lemma 1.1.2 in \cite{Joyce:1996}). Hence the
contribution to the expression for $\eta$ above coming from this
term will cause a positive shift in the effective coupling
constant $k$ for the action $\int_Y *\phi \wedge CS(A^0)$,
reminiscent of the level shift in 3-dimensional Chern-Simons
theory. 

The final contribution to the first Pontrjagin class coming from
$H^4_{\bf 27} (Y,\R )$ is more complicated and generally will not
vanish. Recall it is precisely elements of $H^3_{\bf 27} (Y,\R ) =
H^4_{\bf 27} (Y,\R )$ that parameterize deformations of a given
$G_2$ manifold such that the deformed manifold is also $G_2$.
Hence this contribution would vanish for \lq rigid' $G_2$
manifolds with no deformation moduli (or, of course, for special
$G_2$ manifolds whose first Pontrjagin class has no ${\bf 27}$
part).

The effect on the partition function from the contribution to
$\eta_L$ from $CS_7 (A^0)$ is also rather complicated. We will
simply note that the equations of motion arising from a
modification to the classical action of this kind would be of the
form

\begin{equation}
  \label{}
  *\phi \wedge F = \lambda\, F \wedge F \wedge F \; ,
\end{equation}

\noindent for some constant $\lambda$, which were considered by Leung et al as a deformed version of
Donaldson-Thomas theory.

Just as in 3 dimensions, we expect that the overall $\eta_L (0)$ exponential
prefactor in the partition function will not be a topological invariant. The
task of finding a different regularisation that preserves general covariance is
much more difficult in 7 dimensions and we will not attempt this here.

\subsection{3-cycle worldvolume theory}

Let us now repeat the analysis of the previous section as far as
possible to describe the quantization of the 3-cycle theory. The
effective action for this theory

\begin{equation}
  \begin{split}
    S_{(3)} = \int_M \ep^{abc} \tr \left( A_a \prt_b A_c + \frac{2}{3} A_a A_b A_c + \bt_{ab} D_c f + \frac{1}{2} C_{abc} [ f,f ]
  \right) \\
    +\phi^{aIJ} \tr \left( \theta_I D_a \theta_J + 2\, \bt_{aI} [ \theta_J , f ]
  \right) \; , \\
  \end{split}
\end{equation}

\noindent (derived from OSFT in appendix \ref{app_ghost_3_4_cycle}) is
essentially pure Chern-Simons theory for the gauge field $A_a$ on $M$, which is
a completely solvable theory, plus additional normal mode contributions from
$\theta_I$, whose effect we shall investigate ($D_a = \nabla_a + [A_a,- ]$ on
$M$). It can also be understood as the dimensional reduction of the 7-cycle
action $S_{(7)}$ (after appropriately rescaling $\beta$ and $C$).

In principle a similar modification by normal modes may occur for
open strings ending on special Lagrangian 3-cycles in Calabi-Yau
manifolds in the A-model, though this is not discussed in
\cite{Witten:1992fb}. There is considerable evidence, however,
that the worldvolume theory on a special Lagrangian is essentially
just Chern-Simons theory (up to possible worldsheet instanton
corrections), as this is used, for instance, in open-closed
transitions \cite{Gopakumar:1998ki}. An essential point is that,
aside from $A_a$, none of the other fields in the 3-cycle action
appears at higher than quadratic order in the Lagrangian so they
can be integrated out exactly.

\subsubsection{1-loop partition function}

Let us again simplify matters by quantizing the quadratic part of
the non-linear action $S_{(3)}$, expanded around solutions of the
equations of motion ({\ref{eqn_F_nonflat}}),
({\ref{eqn_gauged_normal}}) for the classical part $S[A,\theta ] =
\int_M CS(A) + \phi^{aIJ} \tr ( \theta_I D_a \theta_J )$ of
$S_{(3)}$.

Expanding $S[A,\theta ]$, for $A_a = A^0_a + B_a$, $\theta_I =
\theta^0_I + \xi_I$, to quadratic order in $(B, \xi )$, around a
classical solution $( A^0 , \theta^0 )$ gives

\begin{equation}
  S [A,\theta ] = S [ A^0 , \theta^0 ] + \int_M \ep^{abc} \tr \left( B_a D_b B_c \right) + \phi^{aIJ} \tr \left( \xi_I D_a \xi_J + \theta^0_I [ B_a , \xi_J ]
  \right)\; ,
\end{equation}

\noindent where $D_a = \nabla_a + [ A^0_a ,-]$. The BV structure of the quadratic
action

\begin{equation}
  S_{cl} [B, \xi ] = \int_M \ep^{abc} \tr ( B_a D_b B_c ) + \phi^{aIJ} \tr
( \xi_I D_a \xi_J + \theta^0_I [ B_a , \xi_J ] ) \; ,
\end{equation}

\noindent is detailed in appendix \ref{app_ghost_3_4_cycle}.  The resulting
gauge-fixed action takes the expected form

\begin{equation}
      S_{cl}[B,\xi ] + \int_M \tr \left( \varphi D_a B^a  + {\bar c}  D_a D^a c
      \right)\; .
\end{equation}

\noindent To compare this quantum theory with Witten's analysis of pure
Chern-Simons theory, let us begin by calculating the contribution to the path
integral from a flat connection (i.e. $A^0$ is flat and $\theta^0 = 0$, solving
({\ref{eqn_F_nonflat}}) and ({\ref{eqn_gauged_normal}})). The modification to
equation (2.8) of \cite{Witten:1988hf} (for the contribution from a flat
connection $A^0$ in pure Chern-Simons theory) due to the normal modes is given
by

\begin{equation}
      \mu ( A^0 , 0) = {\mbox{exp}}(ik S[A^0 ,0]) \; \frac{{\mbox{det}}(D_a D^a)}{\sqrt{{\mbox{det}}(L_- \oplus \phi^{aIJ} D_a
      )}}\; ,
\end{equation}

\noindent where $\phi^{aIJ} D_a$ is understood as a 4x4 antisymmetric matrix of
differential operators. We can go further by making use of the important
identity $\phi^{aIJ} D_a \phi^{bJK} D_b = - \delta^{IK} D_a D^a$, which follows
using $F^0_{ab} =0$. This is related to the fact that $\phi^{aIJ}$, understood
as 3 4x4 matrices, generate an $SU(2)$ subgroup of the $SO(4)$ structure group
of the normal bundle of $M$ and can be understood as Pauli matrices. This allows
us to identify $\phi^{aIJ} D_a$ as a twisted Dirac operator acting on a
4-dimensional vector space, just as Witten did for $L_-$.  Hence going through
the usual Atiyah-Patodi-Singer analysis of the phase factor for the direct sum
of two identical twisted spin complexes over $M \times [0,1]$ (both twisted by
$A^0$) implies the difference of $\eta$-invariants between 2 flat connections $A
= A^0$ and $A=0$ will also be proportional to the pure Chern-Simons action at
$A^0$. Hence this will give essentially the same 1-loop effective action as for
pure Chern-Simons theory except the shift in the level will effectively be
doubled.

Understanding the effect of contributions from more general
solutions of ({\ref{eqn_F_nonflat}}) and
({\ref{eqn_gauged_normal}}) is a more difficult task since not
much is known about this moduli space other than that it contains
flat connections. Formally the contribution from a general
solution $\mu ( A^0 , \theta^0 )$ will be similar to $\mu ( A^0 ,
0 )$ but for replacing $S [ A^0 ,0]$ by $S [ A^0 , \theta^0 ]$ in
the exponential and including an off-block-diagonal component
$\phi^{aIJ} \theta^0_J$ for the determinant in the denominator. It
may prove more convenient to understand such contributions from
the 7-dimensional perspective.

\subsection{4-cycle worldvolume theory}

The action ({\ref{eqn_4cycle_actfull}}) also follows from reduction of the
7-dimensional action $\int_Y *\phi \wedge CS(A)$ on the 4-cycle. The ghost
structure of this theory is derived from OSFT in appendix
\ref{app_ghost_3_4_cycle}, just as for the 3-cycle theory, and again follows
from dimensional reduction of the 7-dimensional theory (up to suitable field
re-scalings) to give the full 4-cycle action

\begin{equation}
  \begin{split}
    S_{(4)} = \int_M \phi^{Iab} \tr \bigl( \tht_I F_{ab} \bigr) +
\frac{2}{3} \, \phi^{IJK} \tr \bigl( \theta_I  \theta_J  \theta_K
\bigr) + \frac{1}{2} \phi^{IJK} \tr \bigl( C_{IJK} [ f,f ] \bigr)  \\
    + 2\, \phi^{Iab} \tr \bigl( \bt_{Ia} D_b f \bigr) + \phi^{IJK} \tr \bigl( \bt_{IJ} [\theta_K ,  f] \bigr) \; . \\
  \end{split}
\end{equation}

\subsubsection{1-loop partition function}

Proceeding as in the previous sections, we quantize the quadratic
part of $S_{(4)}$ by expanding around solutions of
({\ref{eqn_gauged_normal2}}), ({\ref{eqn_FnonSD}}) for the
classical part of $S_{(4)}$

\begin{equation}
  S[A,\theta ] = \int_M \phi^{Iab} \tr \bigl( \tht_I F_{ab} \bigr) + \frac{2}{3}
  \, \phi^{IJK} \tr \bigl( \theta_I \theta_J  \theta_K \bigr) \; .
\end{equation}

\noindent Expanding $S[A,\theta ]$, for $A_a = A^0_a + B_a$, $\theta_I =
\theta^0_I + \xi_I$, to quadratic order in $(B, \xi )$, around a
classical solution $( A^0 , \theta^0 )$ gives

\begin{equation}
  S [A,\theta ] = S [ A^0 , \theta^0 ] + 2\, \int_M \phi^{Iab} \tr \bigl( \xi_I
D_a B_b + \theta^0_I  B_a B_b  \bigr) + \phi^{IJK} \tr \bigl(
\theta^0_I \xi_J \xi_K \bigr) \; ,
\end{equation}

\noindent where $D_a = \nabla_a + [ A^0_a ,-]$. The BV analysis of
the quadratic action 

\begin{equation}
S_{cl} [B, \xi ] = \int_M \phi^{Iab} \tr
\bigl( \xi_I D_a B_b + \theta^0_I  B_a B_b  \bigr) + \phi^{IJK}
\tr \bigl( \theta^0_I \xi_J \xi_K \bigr)
\end{equation}

\noindent is given in appendix \ref{app_ghost_3_4_cycle}, leading to the
expected gauge-fixed action

\begin{equation}
      S_{cl}[B,\xi ] + \int_M \tr \left( \varphi D_a B^a  + {\bar c} D_a D^a c
      \right)\; .
\end{equation}

We will now begin to analyse the quantum structure of this theory
by calculating the contribution to the path integral from an
instanton configuration (i.e. $A^0$ obeys $\phi^{Iab} F_{ab} = 0$
and $\theta^0 = 0$, solving ({\ref{eqn_gauged_normal2}}) and
({\ref{eqn_FnonSD}})). The contribution is given by

\begin{equation}
      \mu ( A^0 , 0) = \frac{{\mbox{det}}(D_a D^a)}{{\mbox{det}}(\phi^{abI} D_b \oplus
      D*)}\; ,
\end{equation}

\noindent where $\phi^{abI} D_b$ is understood as a 4x3 matrix of
differential operators which, together with $D*$ acting on
4-forms, makes up a square 4x4 antisymmetric matrix that provides
an involutive mapping $\Lambda^0 (NM) \oplus \Lambda^4 (M)
\rightarrow \Lambda^1 (M)$. The reason there is no square root in
the denominator is that the differential operator appearing in the
gauge-fixed action is an 8x8 matrix (acting on $B_a$, $\xi_I$ and
$\varphi$) with zeros in the 4x4 block-diagonal entries and the
4x4 operators above in both off-block-diagonal entries. It is not
clear to us if this determinant can be simplified further or
whether it contributes a non-trivial phase factor. The structure
of $\theta^0_I \neq 0$ contributions is also unclear.

%%%%%%%%%%%%%%%%%%%%%%%%%%%%%%%%%%%%%%%%%%%%%%%%%%%%%%%%%%%%%%%%%%%%%%%%%%%%%%%%%%%%%%%%%%%%%

\section{Remarks and open problems}

So far in this paper we have determined the spectrum of the open $G_2$ string
and related it to the worldvolume field theories of branes in a $G_2$ manifold.
In this section we would like to conclude by making some final remarks regarding
issues that still need to be resolved as well as interesting directions for
further research.

\subsection{Holomorphic instantons on special Lagrangians}

In dimensionally reducing the $G_2$ branes on a Calabi-Yau $Z$
times a circle, we have found that we almost reproduce the real
versions of the gauge theories for the open A- and B-models. There
is a discrepancy, however. If one considers a special Lagrangian
$M \sub Z$, with holomorphic open curves $\Sg \sub Z$ ending on
$M$ so that $\prt \Sg \sub M$, then the A-model branes will
receive worldsheet instanton corrections to the standard
Chern-Simons action.  A naive dimensional reduction of the
associative theory on a $G_2$ manifold $Y = Z \times S^1$ gives a
special Lagrangian in $Z$ with the Chern-Simons action without
instanton corrections.

This issue is already present in the closed topological $G_2$
string. When reducing on $CY_3 \times S^1$, the closed $G_2$
string gives a combination of A and B+${\bar {\mbox{B}}}$ models.
But it is non-trivial to see where the worldsheet instanton
corrections in the A-model would come from, given that the $G_2$
theory appears to localize on constant maps. A possible resolution
suggested in \cite{deBoer:2005pt} is that since, unlike a generic
$G_2$ manifold, the manifold $CY_3 \times S^1$ has 2-cycles,
worldsheet instantons may now wrap these 2-cycles. However, upon
closer inspection, this possibility appears rather unlikely. A
much more straightforward explanation is that the worldsheet
instanton contribution is due to topological membranes (i.e.
topological 3-branes of the type discussed in this paper) that
wrap associative cycles of the form $\Sigma \times S^1$ in $CY_3
\times S^1$. Such 3-cycles are indeed associative as long as
$\Sigma$ is a holomorphic curve in the Calabi-Yau manifold.

Returning to the open worldsheet instanton contribution to branes
in the A-model, there are two ways to obtain these from the
topological $G_2$ string on $CY_3 \times S^1$. The first way is to
lift the A-model brane together with the open worldsheet instanton
to a single associative cycle in $CY_3 \times S^1$. This is
similar to the M-theory lift in terms of a single M2-brane of a
configuration of a fundamental string ending on a D2-brane in type
IIA string theory. To describe it, we take a special Lagrangian
3-cycle $C$ in a Calabi-Yau manifold $X$, plus an open holomorphic
curve $\Sigma$. We denote the boundary of $\Sigma$ by
$\gamma\subset C$. We first lift $C$ to $X\times S^1$, which we
describe in terms of a map $C\rightarrow X\times S^1$ which takes
$x\in C$ to $(x,\theta(x)) \in X\times S^1$. Here, $\theta(x)$
describes an $S^1$-valued function on $C$ which we want to have
the property that it winds once around the $S^1$ as we wind once
around the curve $\gamma\subset C$. The lift is therefore
one-to-many, as the image of a point in $\gamma$ is an entire
circle, and because of this the lift of $C$ is an open submanifold
of $X\times S^1$ with boundary $\gamma \times S^1$. We can now
glue the naive lift of $\Sigma$, which is $\Sigma\times S^1$, to
the lift of $C$ to form a closed 3-manifold $M$, since the
boundary of $\Sigma \times S^1$ is also $\gamma \times S^1$. In
this way we have obtained a closed 3-manifold $M\subset X\times
S^1$ which projects down to $C$ and $\Sigma$ upon reduction over
the $S^1$. The 3-manifold $M$ is not calibrated, but we can
compute the integral of $\phi$ over $M$. The result is simply
$\int_C \rho + \int_{\Sigma} k$ if we normalize the size of the
$S^1$ appropriately. The fact that the lift of $C$ winds around
the circle does not yield any additional contribution to $\int_M
\phi$ because the restriction of $k$ to $C$ vanishes identically.

We have thus constructed a closed 3-cycle $M$ such that the integral of $\phi$
over it has the correct structure, geometrically, to yield the worldsheet
instanton contribution. The final step is to minimize the volume of $M$ while
keeping its homology class fixed. This will not change $\int_M \phi$ but
presumably lead to the sought-for associative 3-cycle with the right properties.

In order to push this program further and relate $\int_\Sg k$ to the
(exponentiated) weight of a holomorphic instanton we note that maps $\tht(x)$
which wind about $\gm$ $n$ times will generate contributions such as $n \int_\Sg
k$.  Carefully summing over all lifts of this form with the appropriate weight
might properly reproduce the instanton contributions.

An entirely alternative approach is to lift both $C$ and $\Sigma$
to $C\times S^1$ and $\Sigma \times S^1$. In this way we obtain an
open associative 3-cycle ending on a coassociative 4-cycle in
$X\times S^1$. To analyze whether this makes sense, we consider
the simple example of an open 3-brane in ${\mathbb{R}}^7$
stretched along the 123-direction, ending on a coassociative cycle
stretching in the 2345-direction. If we vary the action
(\ref{eqn_cs_normal_action}) on the 3-brane we obtain a boundary
term

\begin{equation}
S_{\rm boundary} = \int dx^2 dx^3 {\rm tr}(A_3 \delta A_2 -
A_2\delta A_3 + \theta_5 \delta \theta_4 - \theta_4 \delta
\theta_5 +\theta_7 \delta \theta_6 - \theta_6 \delta \theta_7) \;
.
\end{equation}

We obviously want Dirichlet boundary conditions for $\theta_6$ and
$\theta_7$ so that the endpoint of the open 3-brane is confined to
lie in the 4-brane. We also want $\theta_4$ and $\theta_5$ to be
unconstrained at the boundary. If we therefore choose the boundary
condition

\begin{equation}
A_2 = \theta_5 \qquad A_3=\theta_4 \; ,
\end{equation}

the variations all cancel. To preserve these boundary conditions
under a gauge transformation, we need to restrict the gauge
parameter in such a way that its derivatives in the $2,3$ vanish
at the boundary. In this way we indeed find a consistent open
3-brane ending on a 4-brane.

\subsection{Extensions}

The actions we have discovered on topological branes wrapping cycles in a $G_2$
manifold are variants of Chern-Simons theories derived from OSFT.  OSFT itself,
as a generator of perturbative string amplitudes, might need to be augmented by
terms that are locally BRST trivial but none-the-less have global meaning
deriving from the topological structure of the space of string fields.  In the
bosonic open string such questions are currently inaccessible but in the
topological case we see some motivation for local total derivative terms to be
added to the action.  One such potential term is

\begin{equation}\label{eqn_FF_term}
  \int_Y F \wedge F  \wedge \phi
\end{equation}

\noindent that might describe lower dimensional branes dissolved in the seven
dimensional brane.  Such terms might be motivated by analogy with the
Wess-Zumino terms on physical branes.  Note, also, that this reduces to
$F \wedge F \wedge k$ in six-dimensions, a term which appears in the A-model
K\"ahler quantum foam theory \cite{Iqbal:2003ds} which Nekrasov suggests should
be related to holomorphic Chern-Simons theory \cite{Nekrasov:2005bb} (the latter
is, of course, related to our theory by dimensional reduction).  It would be
interesting to try and probe for the existence of such terms directly in the
$G_2$ world-sheet or OSFT theory.

The appearance of the $CS_7(A)$ term in the one-loop partition function suggests
that perhaps this term appears in quantizing the theory and so should have been
included in the original classical action.  

Understanding if such terms do actually appear in these effective actions is
interesting as it may play a role in the conjectured S-duality of the A/B model
topological strings.  In the latter it seems that one may need to consider both
the open and closed theory simultaneously and then terms such as
(\ref{eqn_FF_term}) might play a role in coupling these theories.

\subsection{Relation to twists of super Yang-Mills}

The theories we have found on $G_2$ branes are all topological
theories of the Schwarz type (see \cite{Birmingham:1991ty} for the
terminology) which is no doubt linked to the fact that they are
generated by OSFT.  A similar statement holds for branes in the A-
and B-model.

The worldvolume theory on a brane in a $G_2$ or Calabi-Yau
manifold in a physical model is a twisted, dimensionally reduced
super Yang-Mills (SYM) theory \cite{Bershadsky:1995qy} whose
ground states are topological in nature.  These are related to the
topological field theories that can be constructed by twisting SYM
and considering only the supersymmetric states (by promoting the
twisted supercharge to a BRST operator). Such theories include the
topological action for Donaldson-Witten theory
\cite{Witten:1988ze} as well as its generalizations to higher
dimensions \cite{Baulieu:1997jx}.  These are generally field
theories of the Witten type meaning that the action is itself a
BRST commutator plus a locally trivial term.

Aside from the obvious connection to Chern-Simons theory via OSFT it would be
interesting to understand why the topological theories on branes in topological
string theory are generally of the Schwarz type (which are locally non-trivial)
while the supersymmetric states of the twisted theories on a physical brane can
be studied in a theory that is of the Witten type.

\subsection{Geometric invariants}

One of the most interesting open directions is to investigate the
geometric or topological invariants our open worldvolume gauge
theories compute, and perhaps use them, via open-closed duality,
to discover the connection to the closed topological $G_2$ theory.
It would be interesting to explore the full quantum open string
partition function on  a few examples of $G_2$ manifolds. The
theory on the 3-cycle is basically Chern-Simons theory, while on
the 4-cycle the gauge theory of ASD connections will be related
naturally to Donaldson theory. It would very interesting to find a
role for the partition functions in terms of the full physical
string theory, as well as deepen connections with the mathematics
results in \cite{Leung:2002qs}. Another open problem is to analyze
these invariants in the special case of $CY_3 \times S^1$, and
find a physical understanding of related mathematical invariants
such as the one proposed by Joyce \cite{Joyce:1999tz} counting
special Lagrangian cycles in a Calabi-Yau manifold.

\subsection{Geometric transitions}

Open-closed duality techniques have proven very useful for topological string
theory on Calabi-Yau manifolds. In particular, geometric transitions provide
nice examples where closed topological string amplitudes can be computed from
the gauge theory on the branes, which in this case is just Chern-Simons theory
with possible worldsheet instanton corrections.  Geometric transitions on $G_2$
manifolds in general are less studied, but interesting examples from the full
string theory point of view are exhibited in e.g.
\cite{Gukov:2002zg}\cite{Acharya:1997rh}. In the present
paper we derived the relevant worldvolume gauge theory actions from open
topological strings and so, one of the immediate applications of our results is
to study geometric transitions from the topological $G_2$ string point of view.

\subsection{Mirror symmetry for $G_2$}

Mirror symmetry on a Calabi-Yau 3-fold can be described in terms
of the Strominger-Yau-Zaslow (SYZ) conjecture. One starts with a
special Lagrangian fibration, and then the mirror manifold is
conjectured to be the dual torus fibration over the same base. In
physics language, the action of mirror symmetry on the fibres is
T-duality.  In \cite{Lee:2002fa}, a $G_2$ version of the SYZ
conjecture was suggested, relating coassociative to associative
geometry. Evidence for the $G_2$ mirror symmetry was also found in
$G_2$ compactifications of the physical IIA/IIB string theory on
$G_2$ holonomy manifolds \cite{Acharya:1997rh} \cite{Gukov:2002jv}. It would be
interesting to explore the action of mirror symmetry in the case
of the topological $G_2$ models. A good starting point for this is
by examining automorphisms of the closed $G_2$ string algebra such as those
discussed in \cite{Roiban:2002iv}.

\subsection{Zero Branes}

Although we have not attempted a treatment here it should be possible to reduce
the action (\ref{sevenwv}) to zero dimensions to determine the world-volume of
$D0$-branes on the $G_2$ manifold.  This will be a matrix model which may be
related in an interesting way to the $G_2$ geometry.  

\section*{Acknowledgments}
We would like to thank  Robbert Dijkgraaf, Jos\'{e} Figueroa-O'Farrill, Lotte
Hollands,  Dominic Joyce, Amir-Kian Kashani-Poor, Asad Naqvi,  Nikita Nekrasov,
Martin Ro\v{c}ek, Assaf Shomer, and Erik Verlinde for helpful discussions. The
work of PdM was supported in part by DOE grant DE-FG02-95ER40899 and currently
by a Seggie Brown fellowship.  The work of JdB and SES is supported financially
by the Foundation of Fundamental Research on Matter (FOM).

%%%%%%%%%%%%%%%%%%%%%%%%%%%%
%
% End Main Text
%
%%%%%%%%%%%%%%%%%%%%%%%%%%%%

%%%%%%%%%%%%%%%%%%%%%%%%%%%%
%
% Appendices
%
%%%%%%%%%%%%%%%%%%%%%%%%%%%%

\appendix

%%%%%%%%%%%%%%%%%%%%%%%%%%%%
%
% appendix: Conventions
%
%%%%%%%%%%%%%%%%%%%%%%%%%%%%

\section{Conventions}\label{app_conventions}

In this section we will detail the conventions used in dealing with the
associative 3-form and coassociative 4-form on a 7-manifold with $G_2$
holonomy.  We adopt the conventions of \cite{deBoer:2005pt} since we use many
results from that paper. More details and original references for $G_2$ holonomy
manifolds can be found in that paper.

Although we will generally not have need for the explicit form of $\phi$ or
$*\phi$ we provide a definition in terms of local coordinates, using the
conventions of \cite{deBoer:2005pt}

\begin{align}
  \phi &= \om^{123} + \om^1 \wedge (\om^{45} + \om^{67}) +
    \om^2 \wedge (\om^{46} - \om^{57})
    - \om^3 \wedge (\om^{47} + \om^{56}) \; , \label{eqn_def_phi}\\
  *\phi &= \om^{4567} + \om^{23} \wedge (\om^{67} + \om^{45}) +
        \om^{13} \wedge (\om^{57} - \om^{46}) -
        \om^{12} \wedge (\om^{56} + \om^{47}) \; , \label{eqn_def_sphi}
\end{align}

\noindent where $\om^i$ are vielbeins and $\om^{ij} = \om^i \wedge \om^j$ etc.

We also reproduce some identities for $\phi$ and $*\phi$ from
\cite{deBoer:2005pt} that we will have need of.  The precise factors in these
identities depends on a choice of conventions and normalizations (e.g. their
normalizations are related to those used in \cite{Anguelova:2005cv} by $\phi^{\mbox{\tiny here}}_{\mu\nu\rho} = \frac{1}{3!} \phi^{\mbox{\tiny there}}_{MNP}$ and ${*\phi}^{\mbox{\tiny here}}_{\mu\nu\rho\sigma} = \frac{1}{4!} {*\phi}^{\mbox{\tiny there}}_{MNPQ}$).

\begin{equation}\label{eqn_g2_ident}
  \begin{split}
    \phi^{\mu\al\bt}\phi_{\al\bt\nu} &= \frac{1}{6} \dl^\mu_\nu \; , \\
    (*\phi)_{\mu\nu\al\bt} \phi^{\al\bt\gm} &= \frac{1}{6}
    \phi_{\mu\nu}^{\phnt{\mu\nu}\gm} \; , \\
    \phi_{\mu\nu\gm}\phi^{\gm\al\bt} &= \frac{2}{3}
    (*\phi)_{\mu\nu}^{\phnt{\mu\nu}\al\bt}  + \frac{1}{18} \dl^\al_{[\mu}
    \dl^\bt_{\nu]} \; , \\
    (*\phi)_{\mu\nu\gm\rh} (*\phi)^{\gm\rh\al\bt} &= \frac{1}{12}
    (*\phi)_{\mu\nu}^{\phnt{\mu\nu}\al\bt}  + \frac{1}{72} \dl^\al_{[\mu}
    \dl^\bt_{\nu]} \; .
  \end{split}
\end{equation}

\noindent The exterior algebra on a $G_2$ manifold can be decomposed into
irreducible representations of $G_2$. The decomposition is given as follows

\begin{equation}
  \begin{split}
\Lm^0 = \Lm^0_{\bf 1} \; , &\qquad \Lm^1 = \Lm^1_{\bf 7} \; ,\\
\Lm^2 = \Lm^2_{\bf 7} \oplus \Lm^2_{\bf 14} \; , &\qquad \Lm^3 =
\Lm^3_{\bf 1} \oplus \Lm^3_{\bf 7} \oplus \Lm^3_{\bf 27} \; .
  \end{split}
\end{equation}

\noindent Subscripts here indicate the dimension of the
irreducible representation of $G_2$. The decomposition of higher
degree forms follows by Hodge duality $*\Lm^i_{\bf n} =
\Lm^{7-i}_{\bf n}$.

We will frequently have use for the explicit form of the projectors onto these
representations

\begin{equation}
  \begin{split}
  (\pi^2_{\bf 7})_{\mu\nu}^{\phnt{\mu\nu}\al\bt} &= 6
  \phi_{\mu\nu\gm}\phi^{\gm\al\bt}\; ,
  \\
  (\pi^2_{\bf 14})_{\mu\nu}^{\phnt{\mu\nu}\al\bt} &=
    -4 (*\phi)_{\mu\nu}^{\phnt{\mu\nu}\al\bt}  + \frac{2}{3} \dl^\al_{[\mu}
    \dl^\bt_{\nu]} \; , \\
    (\pi^3_{\bf 1})_{\mu\nu\rh}^{\phnt{\mu\nu\rh}\al\bt\gm} &=
    \frac{1}{7}\phi_{\mu\nu\rh}\phi^{\al\bt\gm} \; .
  \end{split}
\end{equation}

\noindent When a $G_2$ manifold has the structure $CY_3 \times S^1$, there is a
decomposition of $\phi$ and $*\phi$ in terms of $\rho =
{\mbox{Re}}(e^{i\al}\Om)$, $\hrho = {\mbox{Im}}(e^{i\al}\Om)$ and $k$ (where
$\Om$ is the holomorphic 3-form and $k$ is the K\"{a}hler form on $CY_3$). Let $\eta$ be the
volume form on $S^1$ such that $\int_{S^1} \eta = 2\pi R$, then one has the decompositions

\begin{equation}
  \begin{split}
    \phi &= \rho + k \wedge \eta \; ,\\
    *\phi &= \hrho \wedge \eta + \frac{1}{2} k \wedge k \; .
  \end{split}
\end{equation}

\noindent Note that the arbitrary phase $\al$ implies that the real/imaginary
part of $\Om$ is not canonically related to $\phi$ or $*\phi$.  In the paper we
frequently take $\al = 0$ but it is possible to have a $CY_3$ sitting in a $G_2$
with a different alignment of its complex structure.

%%%%%%%%%%%%%%%%%%%%%%%%%%%%
%
% appendix: Ghosts
%
%%%%%%%%%%%%%%%%%%%%%%%%%%%%

\section{Ghost structure}\label{app_ghosts}

A special feature of Chern-Simons theory \cite{Axelrod:1991vq} and
OSFT \cite{Thorn:1986qj} (which have similar functional forms) is
that it is possible to rewrite the gauge-fixed versions of these
theories in the same form as the original theory but with the
gauge or string field replaced by an extended field.

We would now like to argue that the higher (in the sense of
fermion/ghost number) string modes that are BRST closed can be
added to the OSFT action and interpreted as gauge-fixing ghosts or
antifields.  This was suggested in \cite{Witten:1992fb} by noting
that the actions of gauge-fixed CS theory \cite{Axelrod:1991vq}
and OSFT bear a similar form (this is also discussed directly for
physical OSFT in \cite{Thorn:1986qj}).  The main point will be to
re-write gauge-fixed CS theory in terms of a \lq vector
superfield', where the rest of the multiplet comes from the ghosts
and antifields. This superfield has the same expansion as the
string field $\CA(X^\mu, \psi^\mu)$ with the $\psi^\mu_0$'s being
replaced by fermionic coordinates.  In \cite{Axelrod:1991vq} it is
shown that gauge-fixed CS theory written in terms of a superfield
like this has exactly the same action as standard CS theory but
with $A_\mu \ra \CA$.  This allows us to re-interpret
(\ref{eqn_osft_action}) with all the terms in the string field as
a gauge-fixed version of OSFT that reduces to gauge-fixed CS
theory in the large $t$ limit (which would be an exact limit in
topological string theory).

\subsection{7-cycle theory}\label{app_ghost_7cycle}

The normal modes introduce additional complications in gauge
fixing the theory so to avoid these for now we consider first the
theory on the entire $G_2$ manifold.  We will gauge fix this
theory using the Batalin-Vilkovisky (BV) method of quantization
rather than the Faddeev-Popov method, which is used in
\cite{Axelrod:1991vq}, since BV quantization (which is carried out
for OSFT in \cite{Thorn:1986qj}) makes the connection with the
OSFT action (with no constraint on the ghost number of the string
field) more transparent. This connection between closed/open
string field theory and the gauge-fixed
Kodaira-Spencer/holomorphic Chern-Simons description of the
B-model has been established in section 5 of
\cite{Bershadsky:1993cx}.

\subsubsection{BV quantization}

To exhibit this similarity, let us consider the BV quantization of
the classical action (\ref{sevenwv}). We will be brief and refer
the reader to the lecture notes of Henneaux \cite{Henneaux:1989jq}
for more details.

We introduce an anticommuting scalar $c$ corresponding to the BRST
ghost from the gauge symmetry of $S_0 = \int_Y *\phi \wedge
CS(A)$, with associated BRST transformations

\begin{equation}
  \Fs A_\mu = D_\mu c  \; , \qquad \Fs c = \frac{1}{2} [c,c] \; .
\end{equation}

\noindent $A_\mu$ and $c$ have BRST ghost numbers 0 and 1
respectively, and we denote $\Fs$ as the BRST charge associated
with BV quantization (this should not be confused with the BRST
charge $Q$ of the worldsheet or OSFT theory).  Recall that in the
BV formalism, to each field/ghost $\Phi$ one associates an
anti-field/ghost $\Phi^*$ in the Poincar\'{e} dual representation
of the Lorentz group, with opposite Grassmann parity.  Thus we
introduce an anticommuting antifield $A^{*\, \mu}$ for $A_\mu$ and
a commuting antighost $c^*$ for $c$.  These have \lq anti-ghost
number' 1 and 2 respectively, and their BRST transformations are

\begin{equation}
\Fs A^{*\, \mu} = \phi^{\mu\nu\rho} F_{\nu\rho} + [A^{*\, \mu}, c]
\; , \qquad \Fs c^* = D_\mu A^{* \, \mu} +[ c^* ,c] \; .
\end{equation}

\noindent The nilpotent BRST operator $\Fs$ acts on a
doubly-graded complex of functionals, the cohomology of which, in
degree zero (for both gradings) corresponds to gauge-invariant
functionals satisfying the equations of motion. This principle
yields the specific BRST transformations above (see
\cite{Henneaux:1989jq} for more details).  The action of $\Fs$
itself defines a single grading on this complex, given by the
difference of ghost number and antighost number.

An action functional $S$ is defined as the generating function of
the BRST symmetry, such that $\Fs \mathcal{F} = (S, \mathcal{F})$
for any functional $\mathcal{F}$ of the fields or antifields. The
anti-bracket $(-, -)$ is defined as

\begin{equation}
  (A, B) = \frac{\dl^r A}{\dl \Phi} \cdot \frac{\dl^l B}{\dl \Phi^*} -
  \frac{\dl^r A}{\dl \Phi^*} \cdot \frac{\dl^l B}{\dl \Phi} \; .
\end{equation}

\noindent Here $\cdot$ denotes the sum over all fields+ghosts in
$\Phi$, each contracted with their dual antipartners in $\Phi^*$,
the superscripts $r$ and $l$ denote right and left
differentiation. In our case $\Phi = (A_\mu, c)$ and $\Phi^* =
(A^{*\, \mu}, c^*)$. Since the functional $S$ generates the BRST
symmetry it must satisfy $(S, S) = 0$ so that $\Fs^2 = 0$.

These constraints on $S$ allow us to solve for its explicit form,
which is given by an expansion in antighost number (the total
ghost number of the functional must be zero so the terms must have
equal ghost and antighost numbers) of the form

\begin{equation}
  S = S_0 + \int \Phi^* \cdot \Fs \Phi \; .
\end{equation}

\noindent The term $\Phi^* \cdot \Fs \Phi$ again denotes a sum
over all fields and ghosts in $\Phi$, contracted with their
antipartners in $\Phi^*$ (i.e. $A^{*\, \mu} \Fs A_\mu + c^* \Fs c$
in our example). In fact this simple form does not hold in general
but is correct for actions with an irreducible closed gauge
algebra like the one we are considering. For such actions,
Faddeev-Popov gauge-fixing actually suffices but the BV approach
makes the relationship to OSFT more clear.

Thus in our case the generator of the BRST symmetry is

\begin{equation}\label{eqn_action_bv}
  S = \int_Y \phi^{\mu\nu\rh} \, \tr \left(A_\mu \prt_\nu A_\rh +
  \frac{2}{3} A_\mu A_\nu A_\rh + \phi_{\mu\nu\al} A^{*\al} \prt_\rh c +
  \phi_{\mu\nu\al}A^{*\al}[ A_\rh ,c ] +
  \frac{1}{2} \phi_{\mu\nu\rh} c^* [ c , c ] \right) \; . \\
\end{equation}

\noindent It will be clear below that after a relatively trivial
field redefinition, the BV quantized CS theory can be identified
with the OSFT action with unconstrained ghost number string
fields.

It should be noted that this action actually has a larger gauge
symmetry than the original action $S_0$.  This is a standard
feature of BV quantization and is dealt with by restricting the
functional to a graded Lagrangian submanifold of the graded
symplectic manifold spanned by the fields and antifields (this
essentially eliminates the antifield degrees of freedom). In
particular one does not sum over this doubled set of fields in the
path integral. A convenient way to eliminate $\Phi^*$ in terms of
$\Phi$ is via the gauge fermion method whereby one fixes $\Phi^* =
\delta \psi / \delta \Phi$ for some choice of fermionic functional
$\psi [\Phi ]$ of the fields and ghosts only.

\paragraph{Stationary phase expansion.}

Let us not get into the details of gauge-fixing for the full
7-cycle theory above since it will be difficult to evaluate the
exact partition function for this non-linear theory in any case.
Rather, let us consider the theory in the weak coupling limit
where we can restrict to a quadratic expansion about a point of
stationary phase.

The quadratic part of $\int_Y *\phi \wedge CS(A)$, expanded as $A = A^0 + B$
around a solution $A^0$ of the equation of motion $*\phi \wedge F =0$ is given
by

\begin{equation}
  S_{cl} [B] = \int_Y *\phi \wedge \tr ( B \wedge DB ) \; .
\end{equation}

\noindent The BV quantization of this action proceeds as follows.
One finds a minimal solution of the master equations takes the
form

\begin{equation}
  S_{cl}[B] + \int_Y \tr ( B^{*\, \mu} D_\mu c ) \; ,
\end{equation}

\noindent which is invariant under the nilpotent BRST transformations

\begin{equation}
  \Fs B_\mu = D_\mu c \; , \;\; \Fs c = 0 \; , \;\; \Fs B^{* \, \mu} =
  \phi^{\mu\nu\rho} D_\nu B_\rho \; ,
\end{equation}

\noindent involving BRST ghost $c$ and antifield $B^{* \, \mu}$.
Since $c$ is now BRST-invariant, $\Fs c^*$ can be a general
BRST-invariant function. A convenient choice of gauge fermion here
is

\begin{equation}
  \psi = \int_Y \tr ( {\bar c} D^\mu B_\mu ) \; ,
\end{equation}

\noindent in terms of an additional fermionic scalar ${\bar c}$ that is
related to a BRST-trivial bosonic scalar $\varphi$ by $\Fs {\bar c}
= \varphi$. These fields constitute a non-minimal BRST-invariant
addition to the action of the form $\int_Y \tr ( {\bar c}^*
\varphi )$, which still solves the master equation (the antifields
for ${\bar c}$ and $\varphi$ also form a BRST-trivial pair).
Eliminating the antifields via the aforementioned constraint
$\Phi^* = \delta \psi / \delta \Phi$ fixes $B^{*\, \mu} = -D^\mu
{\bar c}$, $c^* =0$, ${\bar c}^* = D^\mu B_\mu$ and $\varphi^*
=0$. Thus the gauge-fixed action takes the familiar form

\begin{equation}
\int_Y *\phi \wedge \tr \left( B \wedge DB \right) + \tr \left(
\varphi D^\mu B_\mu + {\bar c} D^\mu D_\mu c \right) \; ,
\end{equation}

\noindent with $\varphi$ acting as Lagrange multiplier imposing the gauge-fixing
constraint in the action while ${\bar c}$, $c$ correspond to the fermions that
appear in the standard Faddeev-Popov determinant.

\subsubsection{Unconstrained OSFT}

Let us now consider the form of the OSFT action if we remove the
constraint that the string field must be of ghost number one only.
Again, we consider the theory for a brane wrapping the entire
$G_2$ manifold and extend the results of section
\ref{sec_worldvolume_theory} (see e.g. equation
(\ref{eqn_gen_3pt}))

\begin{equation}\label{sevenwvgh}
  \begin{split}
    S_{(7)} &= \int_Y \CA \star Q \CA + \frac{2}{3} \CA \star \CA \star \CA \\
  &= \int_Y \phi^{\mu\nu\rh} \, \tr \left(A_\mu \prt_\nu A_\rh + \frac{2}{3} A_\mu A_\nu A_\rh +
  \bt_{\mu\nu} \prt_\rh f + \bt_{\mu\nu} [ A_\rh ,f ] + \frac{1}{2}
  C_{\mu\nu\rh} \{ f , f \} \right)\\
    &= \int_Y *\phi \wedge \tr \left( A \wedge dA + \frac{2}{3} A \wedge A
	\wedge A  + \bt \wedge D f + \frac{1}{2} C \{ f , f \}
    \right) \; ,
  \end{split}
\end{equation}

\noindent where we have used the full expansion of the string field

\begin{equation}
  \CA(X^\mu, \psi^\mu) = f(X_0) + A_\mu(X_0) \psi_0^\mu + \bt_{\mu\nu}(X_0)
  \psi_0^\mu \psi_0^\nu + C_{\mu\nu\rho}(X_0)\psi_0^\mu \psi_0^\nu
  \psi_0^\rho\; .
\end{equation}

\noindent Here the fields $f \in \Lambda^0_{\bf 1}$, $\beta \in
\Lambda^2_{\bf 7}$ and $C \in \Lambda^3_{\bf 1}$ are respectively
the degree zero, two and three modes of the string field $\CA$ in
the adjoint representation of the gauge group and $D = d +A$ is
the gauge-covariant derivative. The higher degree modes can be
redefined in terms of these lower degree ones via Hodge duality.

The interesting feature of (\ref{sevenwvgh}) is that it has the
same form as the action (\ref{eqn_action_bv}) generated by BV
quantizing the associative Chern-Simons action (\ref{sevenwv}),
corresponding to the ghost number one part of the action.
Specifically, the antifield $A^{*\, \mu}$  in
(\ref{eqn_action_bv}) is identified with the one-form
$\phi^{\mu\nu\rho} \beta_{\nu\rho}$, the antighost $c^*$ is
identified with the zero-form $\phi^{\mu\nu\rho} C_{\mu\nu\rho}$
and the ghost $c$ with $f$.

Note that this identification implies the BV commutator $*A^*
\wedge [A, c]$ must correspond to the OSFT anticommutator $A
\wedge \{f, \bt\}$. As we will explain shortly, this comes about
as a result of the different statistics of $(A^{*\, \mu}, c)$ and
$(\bt_{\mu\nu}, f)$ (which pair of fields appear in the
(anti-)commutator is not so relevant because the cyclicity of the
trace can be used to change them around).

One can check that the linearized equations of motion for $\bt$ and $f$

\begin{equation}
  \begin{split}
	\phi^{\mu\nu\rh} \prt_\rh f(X) = 0 \; ,\\
 \phi^{\mu\nu\rh} \prt_{\rh} \bt_{\mu\nu} (X) = 0 \; ,\\
  \end{split}
\end{equation}

\noindent reproduce the linearized $Q$-closure constraint.  As with the ghost
number one part of the string field, this provides a worldsheet check of the
kinetic terms in the OSFT action.

\paragraph{Ghost correlators.}\label{subsec_ghost_correl}

To check that (\ref{sevenwvgh}) is indeed the correct gauge-fixed
OSFT action, or even effective $D$-brane action, let us calculate
the correlator of the $\bt \wedge A \wedge f$ term on the disc (or
upper half-plane) using the arguments in section
\ref{subsec_3_pt}. We will compare this with the expression for
the 3-pt vertex $*A^* \wedge [A, c]$ in (\ref{eqn_action_bv}). The
subtlety that emerges is that the string correlator will involve
an anticommutator of Grassmann-even fields while the CS 3-pt
function can be recast into a form including a commutator of
Grassmann-odd fields.  This will offset the fact that $\bt$ and
$f$ have different statistics than $A^*$ and $c$.

The twisted correlator

\begin{equation}
  \braket{\bt^i_{\mu\nu}(X)\psi^\mu\psi^\nu A^j_\rh(X) \psi^\rh
  f^k(X)} \; ,
\end{equation}

\noindent receives contributions from the two inequivalent
orderings of the operators on the disk.  The $X$ and $\psi$ CFTs
can be treated separately and indeed the $X$ CFT reduces to an
integral over the $G_2$ manifold (this argument is identical to
the 3-pt function calculation in \cite{deBoer:2005pt}). Using the
$SL(2, \R)$ of the upper half-plane and the cyclicity of the trace
(the above correlator automatically involves a trace over the lie
algebra indices by standard arguments) all possible contributions
will be of the form

\begin{equation}
  (\bt^i_{\mu\nu} t_i A^j_\rh t_j f^k t_k + A_\mu^j t_j \bt^i_{\nu\rh} t_i f^k
  t_k) \psi^\mu \psi^\nu \psi^\rh \; .
\end{equation}

\noindent Here the $t_i$ are a canonically normalized basis of the
lie algebra. As with the $\braket{A A A}$ correlator, the
worldsheet fermions will be contracted with fermions from
$\phi_{\al\bt\gm} \psi^\al \psi^\bt \psi^\gm(z)$ and all the
contractions will actually be equal to each other because the
antisymmetry of the fermions cancels against that of $\phi$ so the
result is some multiple of

\begin{equation}
  \phi^{\mu\nu\rh}\tr[\bt^i_{\mu\nu} t_i A^j_\rh t_j f^k t_k + A_\mu^j t_j
  \bt^i_{\nu\rh} t_i f^k t_k] =
  \phi^{\mu\nu\rh}\tr[A^j_\rh t_j f^k t_k \bt^i_{\mu\nu} t_i + A_\mu^j t_j
  \bt^i_{\nu\rh} t_i f^k t_k] \; .
\end{equation}

\noindent Finally this becomes

\begin{equation}
  \phi^{\mu\nu\rh}A^j_\mu f^k  \bt^i_{\nu\rh} \tr[t_j \{t_k, t_i\}] = *\phi
  \wedge \tr \left( A \wedge \{f, \bt\} \right) \; .
\end{equation}

\noindent Let us now compare this to $*A^* \wedge [A, c]$, using
$\CC_{\mu\nu} = \phi_{\mu\nu\al} A^{*\al}$ for notational
convenience

\begin{equation}
  \begin{split}
  \phi^{\mu\nu\rh} \CC_{\mu\nu}^i  A_\rh^j  c^k \tr[t_i t_j t_k  -  t_i  t_k t_j
  ] &= \phi^{\mu\nu\rh} A_\rh^j \CC_{\mu\nu}^i c^k \tr[t_j t_k t_i  -  t_j  t_i
  t_k ] \\
  &= \phi^{\mu\nu\rh} \tr[- A_\rh^j t_j c^k t_k \CC_{\mu\nu}^i t_i  -  A_\rh^j
  t_j \CC_{\mu\nu}^i t_i  c^k t_k ] \; ,
\end{split}
\end{equation}

\noindent where we have simply used the cyclicity of the trace for
the first equality and the Grassmann-odd nature of the
coefficients $\CC^i$ and $c^k$ for the second.  This then becomes

\begin{equation}
  -{*\phi} \wedge \tr \left( A \wedge \{c, \CC\} \right) \; .
\end{equation}

\noindent Although this seems like a trivial re-writing it is
intended to account for the fact that the statistics of the two
fields are different. Indeed we should perhaps have mapped
gauge-fixed CS theory to string field theory via $\varep \CC \ra
\bt$ and $\varep c \ra f$ with $\varep$ some fixed grassmann-odd
variable.

%%%%%%%%%%%%%%%%%%%%%%%%%%%%%%%%%%%%%%%%%%%%%%%%%%%%%%%%%%%%%%%%%%%%%%%%%%%%%%%%%%%%%%%%%%%%%%%%%%%%%%%%%%%%%%%%%%%%%%%%%%%%%%%%%%%%%%%%%

\subsection{Gauge-fixed OSFT action on calibrated cycles and the BV
formalism}\label{app_ghost_3_4_cycle}

We now consider the form of the OSFT action upon expansion of the
string field on calibrated submanifolds of the $G_2$ manifold, and
how its structure has a natural interpretation in terms of the BV
antifield formalism.  Conceptually this is very similar to the
7-cycle theory but with the added complication of normal modes.

\subsubsection{3-cycle theory}

If one considers the expansion of a general string field on an associative cycle
of a $G_2$ manifold there are many string modes coming from excitations in
the normal directions

\begin{equation}\label{eqn_full_str_fld}
  \CA = f + A_a \psi_0^a +  \tht_I \psi_0^I +
  \bt_{ab} \psi_0^a \psi_0^b +
  \bt_{aI} \psi_0^a \psi_0^I +
  \bt_{IJ} \psi_0^I \psi_0^J +
  C_{abc}\psi_0^a \psi_0^b \psi_0^c + \dots \; .
\end{equation}

\noindent The dots represent the higher modes with at least one normal index in
them.  The expansion above includes all purely tangential modes and the lowest
two orders of normal modes but there are additional higher degree modes with one
or more normal indices which we have not written out.

Including all these contributions, one obtains from OSFT the full
gauge theory action on the 3-cycle

\begin{equation}\label{eqn_deg2_3cycle}
  \begin{split}
        S_{(3)} = \int_M \ep^{abc} \tr \left( A_a \prt_b A_c + \frac{2}{3} A_a A_b A_c + \bt_{ab} D_c f + \frac{1}{2} C_{abc} [ f,f ]
         \right) \\
         +\phi^{aIJ} \tr \left( \theta_I D_a \theta_J + 2\, \bt_{aI} [ \theta_J , f ]
         \right) \; , \\
  \end{split}
\end{equation}

\noindent where we have used the fact that $\beta \in
\Lambda^2_{\bf 7}$ and $C \in \Lambda^3_{\bf 1}$ to derive the
identities $2 \phi^{abc} \beta_{bc} = \phi^{aIJ} \beta_{IJ}$ and
$\frac{2}{7} \phi^{abc} C_{abc} = \phi^{aIJ} C_{aIJ}$ using the
$G_2$ projection operators in appendix A. We have rescaled the
ghost fields for convenience.

The first line in $S_{(3)}$ can be understood in terms of the BV
formalism in exactly the same way we have already described for
$S_{(7)}$. That is, $f$ is the BRST ghost associated with the
gauge symmetry of $A_a$, the antifield $A^{*\, a}$ is identified
with $\ep^{abc} \beta_{bc}$ and the antighost $f^*$ is identified
with $\ep^{abc} C_{abc}$. This would thus lead one to the usual
gauge-fixed action for pure Chern-Simons theory, were it not for
the normal modes. The second line in $S_{(3)}$ would be decoupled,
describing 4 free scalars in the abelian theory. The subtlety this
second line introduces in the non-abelian theory is that it makes
the normal mode action degenerate. In particular it has
non-trivial BRST transformation $\Fs \theta_I = [ \theta_I , f]$
under $\Fs A_a = D_a f$, $\Fs f = \frac{1}{2} [f,f]$ which follows
naturally from the dimensional reduction of the BRST structure in
7 dimensions. Hence $\theta_I$ must also have a fermionic
antifield $\theta^{* \, I}$ which is identified with $\phi^{IJa}
\beta_{Ja}$ in $S_{(3)}$. The corresponding nilpotent BRST
transformations for these antifields

\begin{equation}
  \begin{split}
  \Fs A^{*\, a} &= \ep^{abc} F_{bc} + \phi^{aIJ} [\theta_I , \theta_J
] + [A^{*\, a} ,f] \; ,\\
\Fs \theta^{*\, I} &= 2 \, \phi^{IaJ} D_a \theta_J + [\theta^{*\, I} ,f] \; , \\
\Fs f^* &= D_a A^{* \, a} + [\theta_I , \theta^{*\, I}] +[ f^* ,f]
\; ,
  \end{split}
\end{equation}

\noindent then generate the BRST symmetry of $S_{(3)}$ via the master equation.

The quadratic term in the classical part of $S_{(3)}$, expanded as $A_a = A^0_a
+ B_a$, $\theta_I = \theta^0_I + \xi_I$ around a solution $( A^0_a , \theta^0_I
)$ of the equations of motion is given by

\begin{equation}
  S_{cl} [B, \xi ] = \int_M \ep^{abc} \tr ( B_a D_b B_c ) +
\phi^{aIJ} \tr ( \xi_I D_a \xi_J + \theta^0_I [ B_a , \xi_J ] ) \;
.
\end{equation}

\noindent One finds a minimal solution of the master equations for
this classical action takes the form

\begin{equation}
  S_{cl}[B,\xi ] + \int_M \tr ( B^{*\, a} D_a c + \xi^{*\, I} [ \theta^0_I , c
  ])\; ,
\end{equation}

\noindent which is invariant under the nilpotent BRST transformations

\begin{equation}
   \begin{split}
      \Fs B_a = D_a c \;, \;\; \Fs \xi_I = [ \theta^0_I , c ] \;, \;\; \Fs c =
      0 \;,\\
      \Fs B^{* \, a} = \ep^{abc} D_b B_c + \phi^{aIJ} [ \theta^0_I
      , \xi_J ] \;, \;\; \Fs \xi^{* \, I} &= \phi^{IaJ} ( D_a \xi_J - [
      \theta^0_J , B_a ] ) \; ,
   \end{split}
\end{equation}

\noindent involving BRST ghost $c$ and antifields $B^{* \, a}$, $\xi^{*\,
I}$.

Thus far we have not seen any difference in structure to that one
would get from dimensional reduction of the 7-dimensional theory.
To highlight a potential difference between this reduction and the
quantization of the 3-dimensional theory, considered in its own
right, we make the choice of gauge fermion

\begin{equation}
  \psi = \int_M \tr ( {\bar c} ( D_a B^a + \alpha \, [ \theta^0_I , \xi^I ]
  ))\; ,
\end{equation}

\noindent involving a constant $\alpha$. One has $\alpha =1$ from the
7-dimensional perspective but $\alpha =0$ is a more natural choice
in 3 dimensions. The additional fermionic scalar ${\bar c}$ is
related to a BRST-trivial bosonic scalar $\varphi$ by $\Fs {\bar c}
= \varphi$ as before, and again gives a non-minimal addition to
the action $\int_M \tr ( {\bar c}^* \varphi )$.

Eliminating the antifields via $\Phi^* = \delta \psi / \delta
\Phi$ fixes $B^{*\, a} = -D^a {\bar c}$, $\xi^{*\, I} = - \alpha
\, [ \theta^0_I , {\bar c}]$, $c^* =0$, ${\bar c}^* = D_a B^a +
\alpha \, [ \theta^0_I , \xi^I ]$ and $\varphi^* =0$. Thus the
gauge-fixed action takes the form

\begin{equation}
      S_{cl}[B,\xi ] + \int_M \tr \left( \varphi ( D_a B^a + \alpha\, [ \theta^0_I , \xi^I ] ) + {\bar c} ( D_a D^a c + \alpha \, [ \theta^0_I , [ \theta^{0\, I} , c ] ]
      \right)\; .
\end{equation}

\noindent One can check that the $\alpha$-dependent terms combine to form a
BRST-exact contribution to this action. Thus we argue that any
choice of $\alpha$ will give an equivalent description of the
quantum theory and we will take $\alpha =0$.

\subsubsection{4-cycle theory}

A similar expansion of the string field on a coassociative 4-cycle in the $G_2$ manifold gives rise to the full gauge theory action

\begin{equation}
  \begin{split}
    S_{(4)} = \int_M \phi^{Iab} \tr \bigl( \tht_I F_{ab} \bigr) +
\frac{2}{3} \, \phi^{IJK} \tr \bigl( \theta_I  \theta_J  \theta_K
\bigr) + \frac{1}{2} \phi^{IJK} \tr \bigl( C_{IJK} [ f,f ] \bigr)  \\
    + 2\, \phi^{Iab} \tr \bigl( \bt_{Ia} D_b f \bigr) + \phi^{IJK} \tr \bigl( \bt_{IJ} [\theta_K ,  f] \bigr)  \; ,\\
  \end{split}
\end{equation}

\noindent where $\beta \in \Lambda^2_{\bf 7}$ and $C \in
\Lambda^3_{\bf 1}$ are again used to derive the identities $2
\phi^{IJK} \beta_{JK} = \phi^{Iab} \beta_{ab}$, $\frac{2}{7}
\phi^{IJK} C_{IJK} = \phi^{Iab} C_{Iab}$ for the ghosts on the
4-cycle.

In terms of the BV formalism, $f$ is again the BRST ghost
associated with the gauge symmetry of $A_a$, the antifields
$A^{*\, a}$ and $\theta^{* \,  I}$ are respectively identified
with $\phi^{abI} \beta_{bI}$ and $\phi^{IJK} \beta_{JK}$ whilst
the antighost $f^*$ is identified with $\phi^{IJK} C_{IJK}$. The
BRST transformations of these fields are again

\begin{equation}
  \begin{split}
    \Fs A_a &= D_a f \; , \qquad \Fs \theta_I = [ \theta_I , f ] \; , \qquad
    \Fs f = \frac{1}{2} [f,f] \; , \\
    \Fs A^{*\, a} &= 2\, \phi^{abI} D_b \theta_I + [A^{*\, a} ,f] \; ,\\
    \Fs \theta^{*\, I} &= \phi^{Iab} F_{ab} + \phi^{IJK} [ \theta_J , \theta_K ] + [\theta^{*\, I} ,f] \; ,\\
    \Fs f^* &= D_a A^{* \, a} + [\theta_I , \theta^{*\, I}] +[ f^*
    ,f]\; ,
  \end{split}
\end{equation}

\noindent generating a symmetry of $S_{(4)}$ via the master equation.

The quadratic term in the classical part of $S_{(4)}$, expanded as $A_a = A^0_a
+ B_a$, $\theta_I = \theta^0_I + \xi_I$ around a solution $( A^0_a , \theta^0_I
)$ of the equations of motion is given by

\begin{equation}
  S_{cl} [B, \xi ] = \int_M \phi^{Iab}
\tr \bigl( \xi_I D_a B_b + \theta^0_I  B_a B_b  \bigr) +
\phi^{IJK} \tr \bigl( \theta^0_I \xi_J \xi_K \bigr) \; .
\end{equation}

\noindent One again finds a minimal solution of the master equations for this
classical action takes the form

\begin{equation}
  S_{cl}[B,\xi ] + \int_M \tr ( B^{*\, a} D_a c + \xi^{*\, I} [ \theta^0_I , c
  ])\; ,
\end{equation}

\noindent which is invariant under the nilpotent BRST transformations

\begin{equation}
   \begin{split}
      \Fs B_a = D_a c \; , \;\; \Fs \xi_I = [ \theta^0_I , c ] \; , \;\; \Fs c =
      0 \; ,\\
      \Fs B^{* \, a} = \phi^{abI} \left( D_b \xi_I - [ \theta^0_I
      , B_b ] \right) \; , \;\; \Fs \xi^{* \, I} &= \phi^{Iab} D_a B_b + \phi^{IJK} [
      \theta^0_J , \xi_K ] \; ,
   \end{split}
\end{equation}

\noindent involving BRST ghost $c$ and antifields $B^{* \, a}$, $\xi^{*\, I}$.

We again make the choice of gauge fermion

\begin{equation}
  \psi = \int_M \tr ( {\bar c} ( D_a B^a + \alpha \, [ \theta^0_I , \xi^I ]
  ))\; ,
\end{equation}

\noindent involving the constant $\alpha$. Everything now follows just as
for the 3-cycle case. The gauge-fixed action takes the form

\begin{equation}
      S_{cl}[B,\xi ] + \int_M \tr \left( \varphi ( D_a B^a + \alpha\, [ \theta^0_I , \xi^I ] ) + {\bar c} ( D_a D^a c + \alpha \, [ \theta^0_I , [ \theta^{0\, I} , c ] ]
      \right)\; ,
\end{equation}

\noindent and we choose $\alpha =0$ to ignore the BRST-exact
$\alpha$-dependent terms.

%%%%%%%%%%%%%%%%%%%%%%%%%%%%
%
% appendix: Linearization
%
%%%%%%%%%%%%%%%%%%%%%%%%%%%%

%%%%%%%%%%%%%%%%%%%%%%%%%%%%
%
% appendix: Notes
%
%%%%%%%%%%%%%%%%%%%%%%%%%%%%

%\input{app_notes.tex}

\newpage
\bibliographystyle{utcaps}
\bibliography{master_bib}

\end{document}